\documentclass[fleqn,usenatbib]{mnras}

\usepackage{newtxtext}
\usepackage{newtxmath}

\newcommand{\source}{{Swift~J1727.8-1613}}

\newcommand{\nicer}{NICER}
\newcommand{\hxmt}{\textit{Insight}-HXMT}
\newcommand{\rxte}{\textit{RXTE}}
\newcommand{\sourcem}{{MAXI~J1535-571}}

\usepackage[T1]{fontenc}

\DeclareRobustCommand{\VAN}[3]{#2}
\let\VANthebibliography\thebibliography
\def\thebibliography{\DeclareRobustCommand{\VAN}[3]{##3}\VANthebibliography}

\usepackage{graphicx}	
\usepackage{amsmath}	
\usepackage{gensymb}

\title[Broadband QPO study of Swift~J1727-1613]{A broadband spectral-timing study of QPOs in the bright black hole X-ray binary Swift~J1727.8-1613}
\author[Bollemeijer et al.]{
Niek Bollemeijer,$^{1}$\thanks{E-mail: n.a.bollemeijer@uva.nl}
Phil Uttley$^{1}$
and Bei You$^{2}$\\
$^{1}$Anton Pannekoek Institute for Astronomy, Amsterdam, Science Park 904, NL-1098 NH, The Netherlands\\
$^{2}$School of Physics and Technology, Wuhan University, Wuhan 430072, People's Republic of China\\
}

\date{Accepted XXX. Received YYY; in original form ZZZ}

\pubyear{2024}

\begin{document}
\label{firstpage}
\pagerange{\pageref{firstpage}--\pageref{lastpage}}
\maketitle

\begin{abstract}
\source{} went into outburst in August 2023 and was one of the brightest black hole X-ray binaries (BHXRBs) in recent years, leading to extensive observing campaigns by \nicer{} and \hxmt. The source exhibited strong X-ray variability and showed type-C quasi-periodic oscillations (QPOs) on a wide range of frequencies. The high data quality over a broad range of X-ray energies (0.5-150 keV) enables us to study the energy-dependence of the QPO waveform and the phase lags at the QPO fundamental and second harmonic frequencies. Using the biphase, we find that the QPO waveform is strongly energy-dependent, with energy bands below and above 15-20 keV showing opposite waveform evolution. We interpret the energy-dependence of the waveform as being due to a pivoting spectral component at the second harmonic frequency, with a pivot energy around 15-20 keV. Using the cross-spectrum, we find that the phase lags between energy bands above 7 keV at the QPO fundamental are small, while those at the harmonic frequency are dominated by a separate lag component that extends over a broader range of frequencies and relates to the broadband noise variability. Comparing the energy-dependent results obtained with the bispectrum and the cross-spectrum, we show that these two Fourier products extract different variability components, e.g. the QPO and the broadband noise, at the same frequencies. Finally, we compare \source{} to BHXRB \sourcem{} and find that their spectral-timing properties are similar, indicating that these QPO properties may represent a subset of sources.
\end{abstract}

\begin{keywords}
X-rays: binaries -- black hole physics -- accretion, accretion discs -- X-rays: individual: Swift~J1727.8-1613
\end{keywords}



\section{Introduction}

Low frequency quasi-periodic oscillations (QPOs) are observed in the hard state and hard-intermediate state (HIMS) of many black hole X-ray binaries (BHXRBs, \citealt{Ingram_2019review}). QPOs are strongly associated with X-ray emission linked to a power-law-shaped spectral component coming from an optically thin region close to the black hole, known as the corona \citep{Thorne_1975,Sunyaev_1979}. The nature and geometry of the corona are the subject of an ongoing debate \citep{Done_2007,Poutanen_2018,You_2023} and by placing constraints on the mechanism causing QPOs, we may learn more about the corona as well. Scenarios describing the corona range from a hot inner accretion flow (e.g. \citealt{Ferreira_2006,Veledina_2013hotflow,Kawamura_2022,You_2023}), to an outflow or jet \citep{Markoff_2005,Kylafis_2008,You_2021,Kylafis_2024}. 

The X-ray flux of BHXRBs in the hard state and HIMS does not only vary on the QPO time-scale, but also shows aperiodic, broadband variability. Broadband noise is thought to originate from propagating accretion rate fluctuations in the accretion disc, which has a multicolour blackbody spectrum and dominates at soft X-ray energies \citep{Lyubarskii_1997,Churazov_2001,Arevalo_2006,Ingram_2011,Uttley_2025}. The accretion rate fluctuations cause lags between disc and corona dominated energy bands as they propagate inward, which have been observed in many sources with Fourier-based X-ray spectral-timing techniques \citep{Uttley_2011,De_Marco_2015,De_Marco_2021}. At the same time, differences between coronal cooling and heating due to seed photon variations and accretion rate fluctuations cause the spectral slope of the coronal powerlaw to pivot, leading to lags within the powerlaw component \citep{Kotov_2001,Uttley_2025}. The reverberation of coronal photons on the disc is thought to be another source of lags \citep{Uttley_2014review}. Alternatively, lags may be produced by Comptonisation delays in the corona, where harder photons have been scattered more times within the corona and thus arrive later at the observer \citep{Kazanas_1997,Reig_2003,Bellavita_2022}. 

QPOs show behaviour that is not seen in broadband noise, which has led to the idea that both types of variability have separate origins. For example, many QPO properties, such as the rms variability \citep{Motta_2015}, QPO lags \citep{VandenEijnden_2017} and the phase difference $\Psi$ (see section \ref{sec:qpowaveform}, \citealt{De_Ruiter_2019}) have been observed to depend on observing inclination. These and other findings (see e.g. \citealt{Bollemeijer_2024}) support the class of models that interpret the QPO as a change in the coronal geometry, for example through General-Relativistic Lense-Thirring precession \citep{Ingram_2009LT,Veledina_2013QPO,You_2018}. Evidence for the precession model has been found in the modulation of the Fe K line center with QPO phase due to changing illumination patterns on the disc \citep{Ingram_2016,You_2020}. Other models explain QPOs through naturally arising resonance frequencies between the disc and the corona \citep{Mastichiadis_2022}, corrugation modes \citep{Tsang_2013}, a precessing jet \citep{Ma_2021} or instabilities such as the accretion-ejection instability \citep{Varniere_2002} or the jet instability \citep{Ferreira_2022}. The model of \citet{Garcia_2022} and \citet{Bellavita_2022} focuses on the radiative mechanism behind QPOs to explain the measured time lags between different energy bands at the QPO time-scale.

As a BHXRB transitions through the hard state and HIMS of an outburst, the QPO frequency increases from $\sim$0.1 Hz to $\sim$10 Hz and properties such as the QPO lags and waveform evolve with QPO frequency \citep{Ingram_2019review}. Of the three distinguished types of low frequency QPOs (A, B and C), only type-C QPOs are observed in the hard state and HIMS and we will refer to those simply as QPOs in the current work. The QPO signal is best studied in Fourier space and it is often observed at two frequencies: the largest peak in the power spectrum is by definition called the fundamental, while a smaller peak at twice the fundamental frequency is known as the second harmonic, which we will refer to as simply the harmonic from now on. The Fourier phases of both frequencies are related and can be studied using the bispectrum \citep{Maccarone_2013}. From the phase of the bispectrum (the biphase), the waveform of the QPO can be obtained, which was observed to evolve systematically with QPO frequency in a population study of 14 BHXRBs \citep{De_Ruiter_2019}. 
\newline

\source{} was first detected by the Neil Gehrels Swift Observatory on August 24, 2023 \citep{Kennea_2023,Negoro_2023}. Its luminosity in X-rays increased rapidly over the next few days, reaching a maximum brightness of >7 Crab on August 27 \citep{Palmer_ATel}. With optical spectroscopy of the companion star after the outburst, \citet{MataSanchez_2025} obtained a distance estimate of $3.7\pm0.3$ kpc and a black hole mass of $>3.22 M_{\odot}$, implying that the maximum luminosity was on the order of $\rm{L}_{\rm{Edd}}$. \source{} is probably a medium inclination source, with a strong constraint on the maximum inclination of <74\degree coming from the lack of dipping \citep{MataSanchez_2025}. Spectro-polarimetric modelling attempts yield inclination estimates of 30-50\degree \citep{Svoboda_2024,Peng_2024}.

\source{} is also the first low-mass X-ray binary for which the X-ray polarization was measured in different accretion states, enabling a new view at the nature and geometry of the accretion flow. The polarization measurements suggest that the corona is extended in the plane of the disc during the hard state and HIMS \citep{Veledina_2023,Ingram_2024}, while in the soft state, where the corona is very weak, there is no detectable polarization \citep{Svoboda_2024}. In the dim hard state, the polarization recovers to very similar values as in the bright hard state and HIMS, despite the two orders of magnitude decrease in luminosity \citep{Podgorny_2024}. Radio and OIR observations complement the wealth of X-ray data, making \source{} a powerful laboratory to study accretion in extreme gravity (e.g. \citealt{Baglio_2023,MataSanchez_2023,Vrtilek_2023,Wood_2024}).

We present a thorough X-ray spectral-timing analysis of \source, focusing on QPO properties. We use the extensive sets of observations by \nicer{} and \hxmt{} from August to October 2023, when the source was in the bright hard state and HIMS. \source{} was very bright, has low interstellar absorption ($\rm{N_H}$ $\sim2.4\times10^{21}\rm{ cm}^{-2}$, \citealt{Svoboda_2024}) and showed a slow transition through the successive accretion states, covering almost the full range of QPO frequencies. The nearly daily observations by \nicer{} and \hxmt{} allow an unprecedented view of the QPOs in the hard state and HIMS. In the past decades, many spectral-timing results have been obtained on other BHXRBs using data from the Rossi X-ray Timing Explorer (\rxte), which had an effective energy range of 2-60 keV. With \nicer{} and \hxmt{} we can extend that energy range to 0.5-150 keV, repeat analyses and compare results to other sources that were observed by both telescopes to gain more understanding of the \rxte{} results. 

The paper is structured as follows. In Section \ref{sec:data}, we describe how we obtained and reprocessed the data, extracting usable data despite the problems associated with the light leak in \nicer. In Section \ref{sec:qpowaveform} and \ref{subsec:qpolag}, we study the QPO evolution by measuring the QPO waveform and QPO lags for different energy bands. Because the energy-dependence of some of the spectral-timing properties was not measured in such detail before, we compare our results for \source{} to the smaller data set for \sourcem{} and find that both sources show comparable behaviour, implying that \source{} is not unique and may represent at least a subset of BHXRBs. In Section \ref{sec:discussion}, we give a brief interpretation of our results and suggest paths for further research.

\section{Data}
\label{sec:data}
\subsection{NICER}
\label{sec:data_nicer}

\begin{figure}
    \centering
    \includegraphics[width=\linewidth]{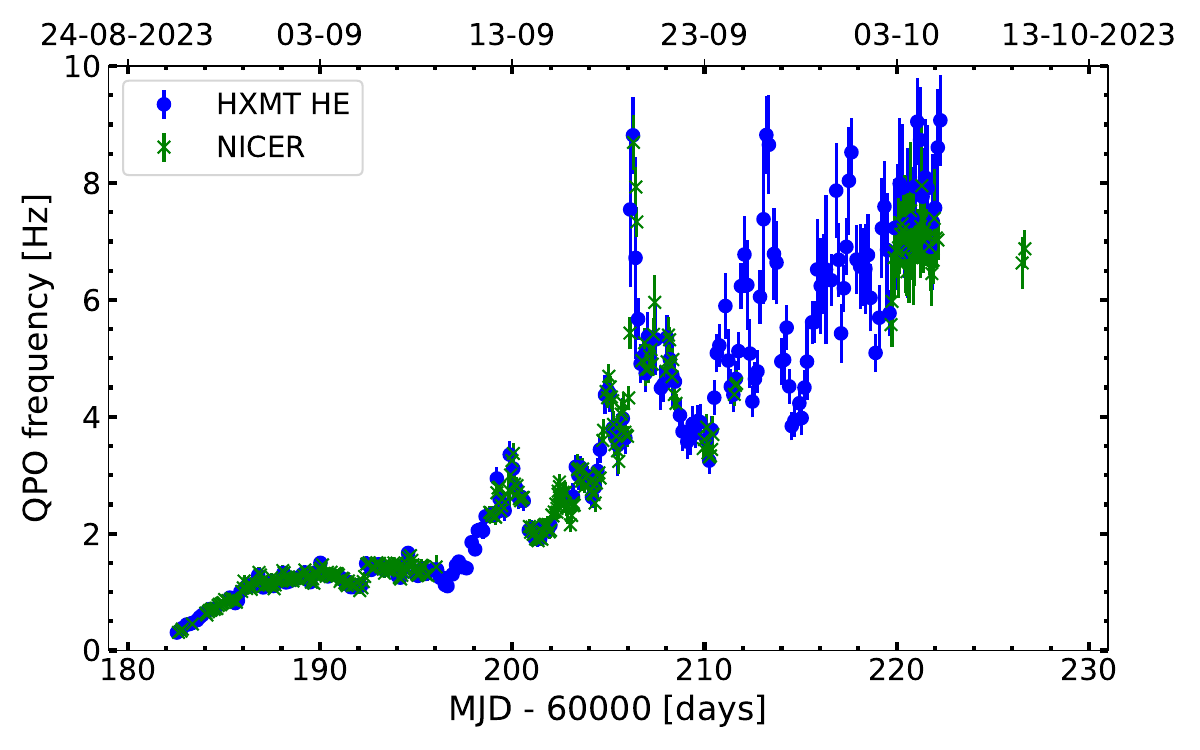}
    \caption{The QPO frequency as a function of time in \source{} for all data used as measured in the 0.5-10 keV band with \nicer{} (green) and the 28-100 keV band with the HE instrument of \hxmt{} (blue). The upper x-axis shows the start of each date corresponding to the MJD value on the lower x-axis. The error bars represent the FWHMs of the Lorentzian fitted to the power spectrum.}
    \label{fig:QPOf_vs_MJD}
\end{figure}

\nicer{} \citep{Gendreau_2016} is an X-ray telescope located on the International Space Station (ISS) and has been in operation since 2017. Its large effective area at low X-ray energies (0.3-12 keV) and high count rate capability has revealed many new properties of BHXRBs. \nicer{} started an intensive observing campaign of \source{} on August 25 \citep{Oconnor_2023}, monitoring the source closely until October 9, when \source{} left its visibility window due to solar constraints. Due to technical constraints, there were two other observation gaps and no observations were made on September 9 and 10 and between September 23 and October 2, as is visible in Fig. \ref{fig:QPOf_vs_MJD}. 

We used the \texttt{nicerl2} pipeline of HEASoft (v6.33) to reprocess the data \citep{FTOOLS_2014}. However, for many observations, standard data reduction and reprocessing tasks result in no Good Time Intervals (GTIs), complicating data usage and extraction. To understand the caveats of working with NICER's observations of \source, we will summarise relevant instrument properties and outline some of the problems that are associated with this particular data set.

NICER consists of 7 Measurement and Power Units (MPUs), each containing 8 Focal Plane Modules (FPMs). Data from the FPMs can be analysed individually. Of the 56 initial FPMs, 52 have been functional since the launch of the telescope. Although NICER is a powerful timing instrument, with small deadtime and high timing resolution, high X-ray source count rates in a given MPU can lead to internal telemetry saturation, resulting in data loss and fragmented GTIs, heavily impacting the shape of the light curve and any derived spectral-timing property. By switching off a number of FPMs, the total observed count rate drops, solving the telemetry problem, at the cost of a reduced number of observed photons. For a very bright source like \source, a decreased count rate is generally not a large problem. In the analysed data set, the number of FPMs used ranges between 3 and 21. When we calculate spectral-timing properties as a function of QPO frequency, we combine data from different observations, that may have different numbers of FPMs switched on. As such, the measurement of the spectral-timing property will be affected more by light curves with large numbers of FPMs switched on than those with smaller numbers of FPMs. We tried normalising all light curves used by the number of FPMs, which does not change the results significantly and mainly has the effect of increasing the weight of noise in observations with a lower number of detectors.

Another challenge is posed by sunlight leaking into NICER, after a puncture of the thermal film protecting the telescope detectors from stray light by a piece of space debris on May 22, 2023\footnote{\url{https://heasarc.gsfc.nasa.gov/docs/nicer/analysis_threads/light-leak-overview/}}. The increased amount of sunlight reaching the X-ray detectors leads to an additional source of charge, prompting the detectors to reset more often. Although such resets, known as undershoots, occur in any functional detector, they can affect the energy scale of the telescope when their number is very high (>500 undershoots/s). Also, high undershoot rates cause an extended noise peak at low energies (affecting up to $\lesssim$0.6 keV) and lead to an increase in detector deadtime.\footnote{\url{https://heasarc.gsfc.nasa.gov/docs/nicer/analysis_threads/undershoot-intro/}}

By running \texttt{nicerl2} with adjusted settings, allowing for higher undershoot rates, and by excluding detectors with very high undershoot rates or other problematic events such as data packet losses, we extracted a usable set of observations. The choice of the \texttt{nicerl2} settings, especially the \texttt{underonly\_range} parameter, are discussed in more detail in Appendix \ref{sec:app_nicerdata}. The change in energy scale and soft X-ray noise associated with the undershoot rates will cause reduced source count rates and increased noise levels at low energies. We have determined the effects of these changes to be small when doing broadband spectral-timing analysis. In Appendix \ref{sec:app_lightleak}, we compare orbit day and night observations to show that the effects of the light leak are limited for spectral-timing analyses, but they may be very significant for spectroscopic analysis.

Despite \nicer's modest deadtime, the very high count rates associated with \source{} are still affected significantly and we apply the same methods we used in \citet{Ingram_2024} to mitigate the effects of deadtime. When calculating the cross-spectrum, we use pairs of light curves made with separate sets of FPMs. If we were to use light curves from the same detectors, the significant deadtime introduces an anti-correlation between energy bands at high frequencies, as a photon observed in one energy band will prevent another photon being detected for a short while. Deadtime also decreases the Poisson noise level, causing it to deviate from $2/\langle x \rangle$ (in fractional rms units), where $\langle x \rangle$ is the count rate \citep{Uttley_2014review}. To obtain the noise level, we measured the power between 200 and 250 Hz (using a 0.002 s time resolution), where there is no significant source signal. To verify that the obtained power spectrum is correct, we also calculated the cospectrum, the real part of the cross-spectrum, of lightcurves from two identical energy bands from independent detectors, which can be used as an equivalent when there is significant deadtime \citep{Bachetti_2015}. All cross-spectral parameters, such as the lags and the coherence, may also be affected by deadtime or telemetry problems within a single MPU. We make sure that the two light curves for a single energy band are composed of data from different FPMs as well as different MPUs, to limit the possibility of any correlated systematic signal. The main downside of using separate sets of detectors for cross-spectral measurements is that the count rate in each light curve is decreased by a factor $\sim$2, but given the high count rates from \source, the advantages outweigh the larger error bars.

\begin{figure*}
    \centering
    \includegraphics[width=175mm]{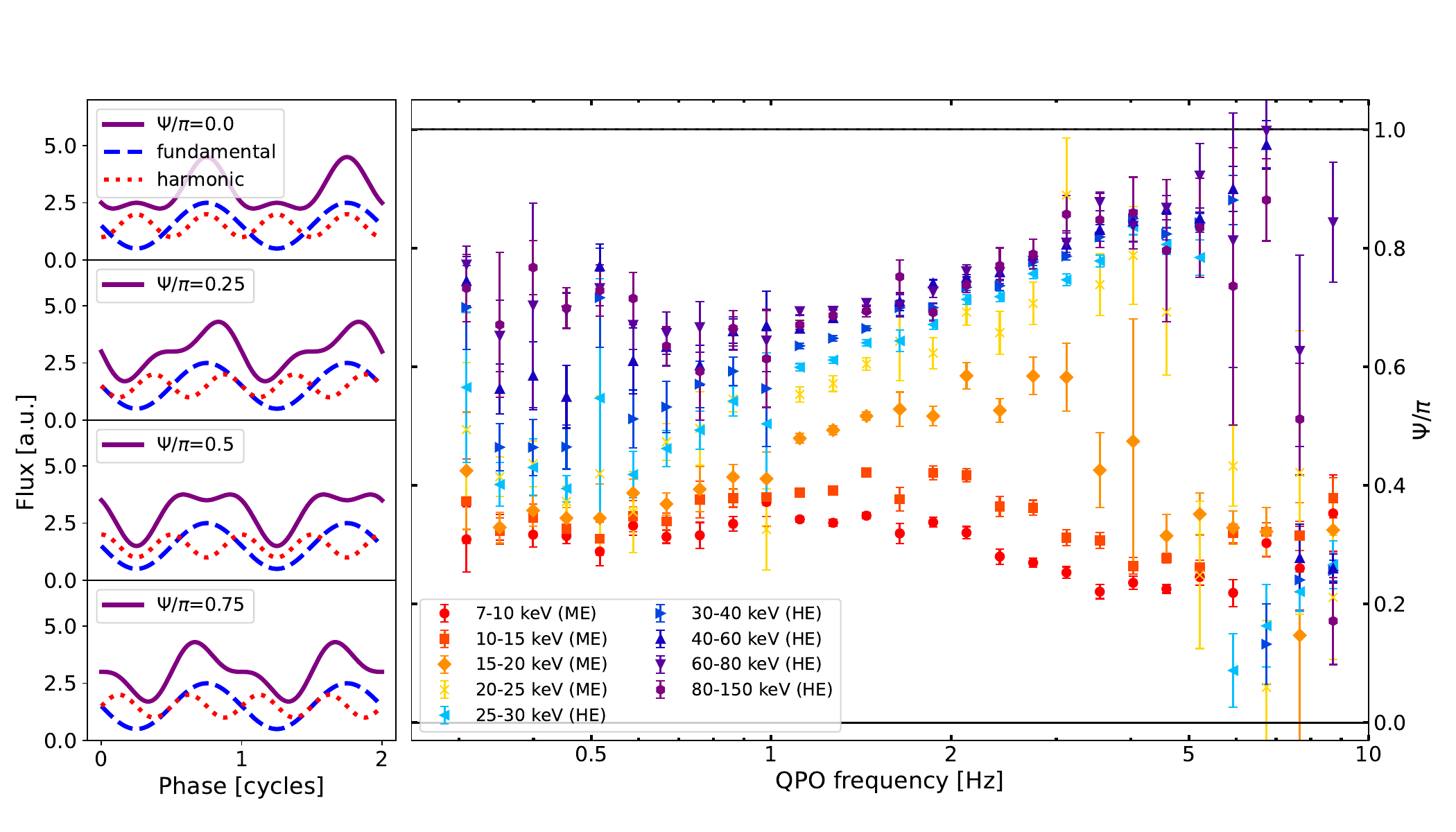}
    \caption{The main panel on the right shows phase difference $\Psi$ in units of cycles for \source{} calculated from \hxmt{} data, using the biphase at the QPO fundamental and harmonic frequencies for different QPO frequencies and energy bands. The energy bands below $\sim$15 keV show a $\Psi$ evolution that is largely consistent with the results of \citet{De_Ruiter_2019} for high-inclination BHXRBs, where similar energies were used. Above $\sim$20 keV, the evolution of $\Psi$ is in the opposite direction, especially above $\sim$2 Hz. The four panels on the left show example waveforms for four different values of $\Psi$, with the harmonic amplitude being half the fundamental amplitude.
    }
    \label{fig:Psi_J1727}
\end{figure*}
\begin{figure}
    \centering
    \includegraphics[width=\linewidth]{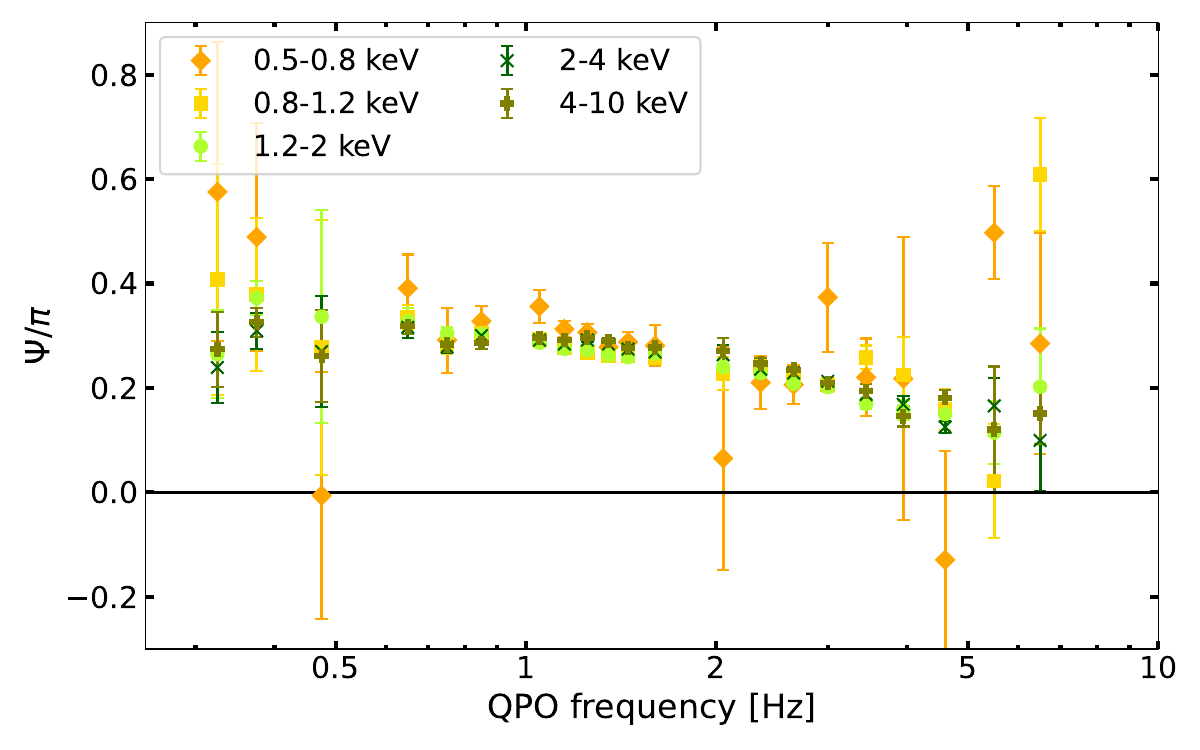}
    \caption{The phase difference $\Psi$ for \source{} from \nicer{} data, using the biphase at the QPO fundamental and harmonic frequencies for different QPO frequencies and energy bands. All energy bands show similar behaviour, decreasing from $\Psi\sim0.3\pi$ rad at low QPO frequencies to $\sim0$ around 6 Hz. At high QPO frequencies and for the softest energy band, the error bars and scatter are large due to small numbers of usable detectors and sparsity of reliable observations. The behaviour of $\Psi$ is consistent with the measurements by \citet{De_Ruiter_2019} and similar to the results for \hxmt{} data up to $\sim$15 keV, as shown in Fig. \ref{fig:Psi_J1727}. For plotting clarity, we wrapped values above 0.8 around by subtracting 1 cycle. 
    }
    \label{fig:Psi_J1727NICER}
\end{figure}

\subsection{\textit{Insight}-HXMT}
\label{sec:data_hxmt}
The Hard X-ray Modulation Telescope (\hxmt, \citealt{Zhang_2014}) performed an extensive observing campaign of \source{} during its hard and hard-intermediate state, observing the source almost every day starting on August 25 and ending on October 6 due to solar constraints. 

\hxmt{} consists of three different instruments, covering a wide range of X-ray energies. In the current work, we use the Medium Energy (ME) instrument for the 7-25 keV range and the High Energy (HE) instrument for the 25-150 keV energy range of all ObsIDs ranging from P0614338001 to P0614338035. We do not use LE data, because the 2-10 keV energy range is already covered by \nicer. All \hxmt{} data used are publicly available online\footnote{\url{http://archive.hxmt.cn/proposal}}.. The data were then reprocessed with the HXMT Data Analysis Software (HXMTDAS, version 2.06) \texttt{hpipeline} function, using the default settings. To create the light curves analysed in the current work, we binned the screened event files by their arrival time and measured energy (the \texttt{TIME} and \texttt{PI} columns, respectively). We did not subtract background light curves. For the HE instrument, the background is expected to vary only on longer timescales ($\gtrsim$100 s, \citealt{Liao_2020HE}). The background light curve is determined by averaging over 16 s 0.05intervals by the background modelling function \texttt{hebkgmap} and interpolating between those data points. Similary, the background light curves of ME show variations only on a timescale of 100s of seconds \citep{Guo_2024}. Subtracting those background light curves would aid in estimating the fractional rms, which is artificially reduced when background is included \citep{Zhou_2022_cygx1}, but would otherwise not change the cross-spectral measurements, as there is no correlated variability in the background on timescales of seconds. To make sure that the background does not influence our measurements, we compared spectral-timing results with and without background subtraction and found that only on long timescales ($\lesssim$0.1 Hz) do we observe any change, which is still well within the error bars. In Appendix \ref{sec:app_background}, we show the effects of subtracting background for the latest observation included in this paper, P0614338035.

At the high source count rates from \source, deadtime can significantly alter spectral-timing measurements with \hxmt. Deadtime is especially relevant in the ME instrument, where the average deadtime is on the order of 250 $\rm{\mu s}$ \citep{Tuo_2024}. To minimise the effects of deadtime, we follow the same procedure as described for \nicer{} and use pairs of light curves from separate sets of the 9 Field Programmable Gate Arrays (FPGAs) in ME to calculate the cross-spectrum and any derived quantities (see Sections \ref{sec:qpowaveform} and \ref{subsec:qpolag} for details). When calculating the cross-spectrum between light curves from ME and HE, the instruments are independent and we use the full set of detectors in both instruments. Finally, when using HE to calculate the bispectrum and test the noise correction (see Section \ref{sec:qpowaveform}), we use the first 6 of the 18 NaI (Tl)/CsI (Na) detectors and the final 12 to make pairs of independent light curves, since some deadtime is shared between sets of 6 detectors \citep{Xiao_2020}. Similar to the procedure for \nicer{} data, we estimate the Poisson noise level by fitting a constant to the 200-250 Hz frequency range of the power spectrum.

\section{Analysis and results}
\label{sec:analysis_results}
In this section, we show the results of the different analyses we performed on the data. In subsection \ref{sec:qpowaveform} and \ref{subsec:qpolag}, we show the QPO evolution during the outburst by measuring the energy-dependent waveform and phase lags, respectively. Because the QPO frequency varies back and forth during the outburst, we combined snapshots of one orbit (for \nicer) and sub-observations, consisting of a few orbits and defined by the last two numbers of the full observation ID (for \hxmt) by their QPO frequency. To obtain the QPO frequency for each of those, we fitted Lorentzians to power spectra of light curves from energy ranges 0.5-10 keV (\nicer) or 28-100 keV (HE on \hxmt). We generally used 64 s segments for both telescopes, except for the observations of \nicer{} in October 2023. There, the ($>5$ Hz) QPO is less strong and to ensure we have sufficient data to be able to determine the QPO frequency, we used 16 s segments. We created power spectra for each snapshot for \nicer{} and each sub-observation for \hxmt. Some snapshots or sub-observations did not show any QPOs and we excluded them from our analysis. By fitting Lorentzians to the power spectra, we determined the QPO frequency of each snapshot (\nicer) or sub-observation (\hxmt), as is shown in Fig. \ref{fig:QPOf_vs_MJD} for all observations used. The great coverage of \source{} is clearly visible, as well as the fast changes in the QPO frequency starting around MJD 60198, which is when the flare state begins \citep{Yu_2024}. For our spectral-timing analysis, we grouped QPO frequencies in geometrically spaced frequency bins, as is visible in e.g. Fig. \ref{fig:Psi_J1727}. 

\subsection{QPO waveform}
\label{sec:qpowaveform}

\source{} shows very strong QPOs during its outburst and we will study their waveform and lag properties in the following two sections. We assume here that the QPO signal has an underlying non-sinusoidal waveform, which can be parametrised in terms of harmonics when using Fast Fourier transforms (FFTs), as was shown by e.g. \citet{Ingram_2014} and \citet{De_Ruiter_2019}. We note that the QPO waveform can also be studied with Hilbert-Huang transforms (HHTs), as was demonstrated by \citet{Shui_2023} and \citet{Shui_2024}, albeit with suppression of the harmonic content in the waveform. \citet{Shui_2023} and \citet{Shui_2024} compared FFT and HHT-based methods and found that both returned consistent results in terms of the overall shape of the waveform. Recently, \citet{Yan_2024} showed that with a different HHT-based decomposition technique, it is possible to study non-linearities in X-ray light curves, which leads to different contributions from harmonic signals. Although we believe the use of HHT-based analyses is worth further study, we only use FFT-based methods in the current work, as they allow us to compare our results to previous findings, such as those by \citet{De_Ruiter_2019}. In Appendix \ref{sec:app_simulations}, we show with simulated data that our methods can reliably reproduce the QPO waveform.

The QPO waveform is defined to first order by the amplitude ratio of the harmonic and fundamental signal of the QPO and the phase difference $\Psi$ between the fundamental and harmonic \citep{Ingram_2014}. \citet{De_Ruiter_2019} found that the waveform depends on QPO frequency and on the observed binary inclination. The authors interpret the inclination dependence of the waveform evolution as a signature of the geometric origin of QPOs, as various beaming and solid angle effects can cause such a relation to emerge \citep{You_2018}. $\Psi$ follows a clear frequency-dependent evolution for (almost) all high inclination sources, declining smoothly from $\sim$0.3$\pi$ rad at QPO frequencies below 1 Hz to $\sim$0 when the QPO frequency exceeds 5 Hz. No single trend can be discerned for low inclination sources. 

We measured the harmonic-fundamental phase difference $\Psi$ for \source, which has excellent \hxmt{} and \nicer{} coverage of all QPO frequencies between 0.3 and 10 Hz. Where \citet{De_Ruiter_2019} used a 2-13 keV RXTE band and found no energy dependence when using energy bands of 2-7 and 7-13 keV, we broadened our view to multiple energy bands between 0.5 and 150 keV. We measured phase difference $\Psi$ for different observations grouped by their QPO frequency, following the approach of \citet{Nathan_2022} and using the biphase. The biphase is the phase of the bispectrum, defined by 
\begin{equation}
\label{eq:bispectrum}
    B(\nu_{12}) = F(\nu_1)F(\nu_2)F^{\ast}(\nu_1+\nu_2),
\end{equation}
where $B(\nu_{ij})$ is the bispectrum, $F(\nu_i)$ is the FFT of the light curve at frequency $\nu_i$ and $F^\ast(\nu_i)$ is its complex conjugate. The bispectrum measures phase-correlations between Fourier frequencies in a single signal. If $\nu_1=\nu_2$, the outcome of Equation \ref{eq:bispectrum} is called the autobispectrum and it contains information about the relation between frequency $\nu_1$ and double that frequency, like the fundamental and harmonic of a QPO. A biphase of zero indicates that the fundamental and harmonic are in phase with each other. Because Poisson noise can affect the measured biphase \citep{Uttley_2005,Maccarone_2013}, we apply the formula to correct for the noise in Appendix A of \citet{Nathan_2022}, which was obtained from \citet{Wirnitzer_85} and can be written as
\begin{equation}
    \begin{aligned}
        B_m(\nu) = &{} F_m(\nu)F_m(\nu)F^\ast_m(2\nu)\\
        &-2|F_m(\nu)F^\ast_m(\nu)|^2 -|F_m(2\nu)F^\ast_m(2\nu)|^2
        +2N_m.
    \end{aligned}
\label{eq:bispectrum_noisecorrection}
\end{equation}
In this expression, $F_m(\nu)$ is the FFT of a given segment and $N_m$ is the number of photons in that same light curve segment. We also measured the biphase using separate sets of telescope detectors for $F_m(\nu)$ and $F_m(2\nu)$ to remove any correlations between different frequencies due to Poisson noise or deadtime, which returned very similar results, but with larger error bars. We therefore only show the results while using the full available set of detectors and correcting for Poisson noise with equation \ref{eq:bispectrum_noisecorrection}. 

The bispectrum is complex-valued and its phase can be converted to the phase difference $\Psi$ through $\Psi=\frac{1}{2}\rm{arg}(B(\nu_{12}))$. Following the approach of \citet{Nathan_2022}, we calculate the biphase for light curves of $\sim8$ times the QPO period, which is comparable to the typical coherent interval length of the QPO, related to the width of the QPO peak \citep{Ingram_2015,VandenEijnden_2016origin}. We divided the obtained value of $\Psi$ by $\pi$, such that it ranges between 0 and 1, to enable comparison to Fig. 5 in \citet{De_Ruiter_2019}. $\Psi/\pi$ is a cyclic quantity, so 0 and 1 have exactly the same meaning.

In Figs. \ref{fig:Psi_J1727} and \ref{fig:Psi_J1727NICER}, we show how the phase difference between the fundamental and harmonic changes with QPO frequency, which was observed to be the case for many BHXRBs by \citet{De_Ruiter_2019}. However, \source{} transitioned through different QPO frequencies relatively slowly, especially below $\sim$3 Hz (see Fig. \ref{fig:QPOf_vs_MJD}), and was so bright that high quality light curves could be obtained for multiple energy bands with \hxmt{} for QPO frequencies in the 0.3-10 Hz range. The result is shown in Fig. \ref{fig:Psi_J1727}, where we see two broad trends. The energy bands up to $\sim15$ keV have $\Psi\sim0.3\pi$ rad for all QPO frequencies, which is similar to the results for high-inclination sources presented by \citet{De_Ruiter_2019} at low frequencies. At high frequencies, the sources in \citet{De_Ruiter_2019} have $\Psi\sim0$ rad, while we measure values of $\Psi$ between 0.2 and 0.4 $\pi$ rad. For higher energies, $\Psi$ is generally >0.5 $\pi$ rad and increases with QPO frequency, until $\Psi$ is is phase-wrapped and becomes $\sim0.3\pi$ rad (or highly uncertain) for all energies at a QPO frequency of 8 Hz. We used \nicer{} to extend our analysis to low energies, as shown in Fig. \ref{fig:Psi_J1727NICER}. Remarkably, the phase difference $\Psi$ is very similar for all energies between 0.5 and 10 keV and the general trend in the \nicer{} data is consistent with the measurements in energy bands below 15 keV in \hxmt{} data, although the data did not allow measurement of $\Psi$ for the highest QPO frequencies.

\begin{figure*}
    \centering
    \includegraphics[width=175mm]{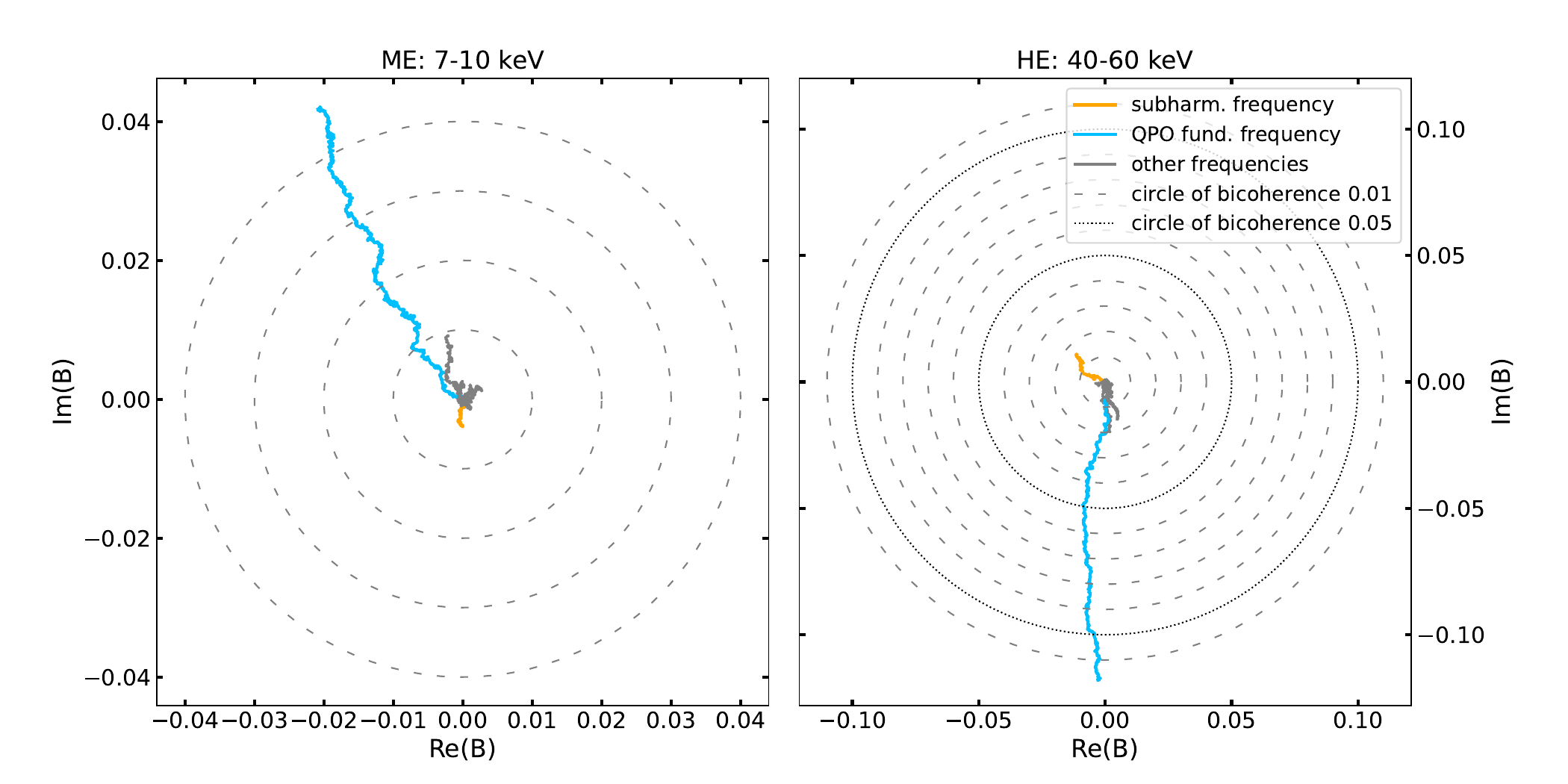}
    \caption{The `jellyfish plots' of the bispectrum of two energy bands (7-10 and 40-60 keV) for \hxmt{} data with a QPO frequency around 2.1 Hz. The real and imaginary parts for the biphase of each light curve segment are shown for different frequencies, with the QPO frequency bispectra shown in blue. The axes are normalised such that the total amplitude shows the bicoherence. The bispectra of the subharmonic and fundamental frequencies are shown in orange and have a distinct biphase. All other non-QPO frequencies are shown in grey. It is clear that the biphase of the fundamental and harmonic frequency, shown in blue, has the highest bicoherence and that the biphase is shifted by about half a cycle between both energy bands. In the right panel, the smaller, grey line in a similar direction as the QPO frequency corresponds to a frequency close to the QPO frequency and thus has increased bicoherence and similar biphase.}
    \label{fig:jellyfish}
\end{figure*}

To emphasise the difference between the waveform of the energy bands below and above 15-20 keV in \hxmt{} data, we present two `jellyfish plots' as introduced by \citet{Nathan_2022} in Fig. \ref{fig:jellyfish}. The two panels in Fig. \ref{fig:jellyfish} show the cumulative sum of the real and imaginary parts of the bispectrum for different frequencies, normalised such that the final amplitude of the plotted lines represents the bicoherence. The bicoherence is a measure of the strength of the connection between different Fourier frequencies \citep{Uttley_2005,Arur_2019,Arur_2020}. The two panels in Fig. \ref{fig:jellyfish} show the real and imaginary parts of the bispectrum for two different energy bands, 7-10 keV in the left and 40-60 keV in the right panel. The bispectrum of the QPO fundamental and harmonic is shown in blue, while the bispectrum of the subharmonic and fundamental is orange. The grey lines correspond to all other pairs of frequencies. We find that the subharmonic biphase has its own waveform, which is much weaker than the biphase of the fundamental and harmonic and may be connected to the `hypotenuse pattern' observed in other sources \citep{Arur_2019,Arur_2020}. We leave the origin of the deviating subharmonic biphase for future work.

We determined error bars on $\Psi$ with bootstrapping in the following way. For a data set with $N$ pairs of light curve segments with length $\sim$8 QPO cycles, we drew, with replacement, $N$ light curve segments from the full set of segments with a similar QPO frequency. We then calculated $\Psi$ using the method explained above and saved the outcome. After 100 bootstrap samples were generated, we created a histogram with the obtained $\Psi$ values, the standard deviation of which is the error on the measurement. From the size of the error bars in Fig. \ref{fig:Psi_J1727}, it is obvious that the differences between low and high energies are very significant, at least between QPO frequencies of 1 and 6 Hz. Although an energy-dependence of the waveform is expected, based on the QPO-phase-dependence of spectral parameters observed through QPO-phase-resolved spectroscopy \citep{Ingram_2016,Nathan_2022}, the dramatic bifurcation of $\Psi$ for different energy bands that we observe in the current work is not. We will discuss a brief interpretation of this new result in Section \ref{sec:discussion_waveform}.

\begin{figure}
    \centering
    \includegraphics[width=\linewidth]{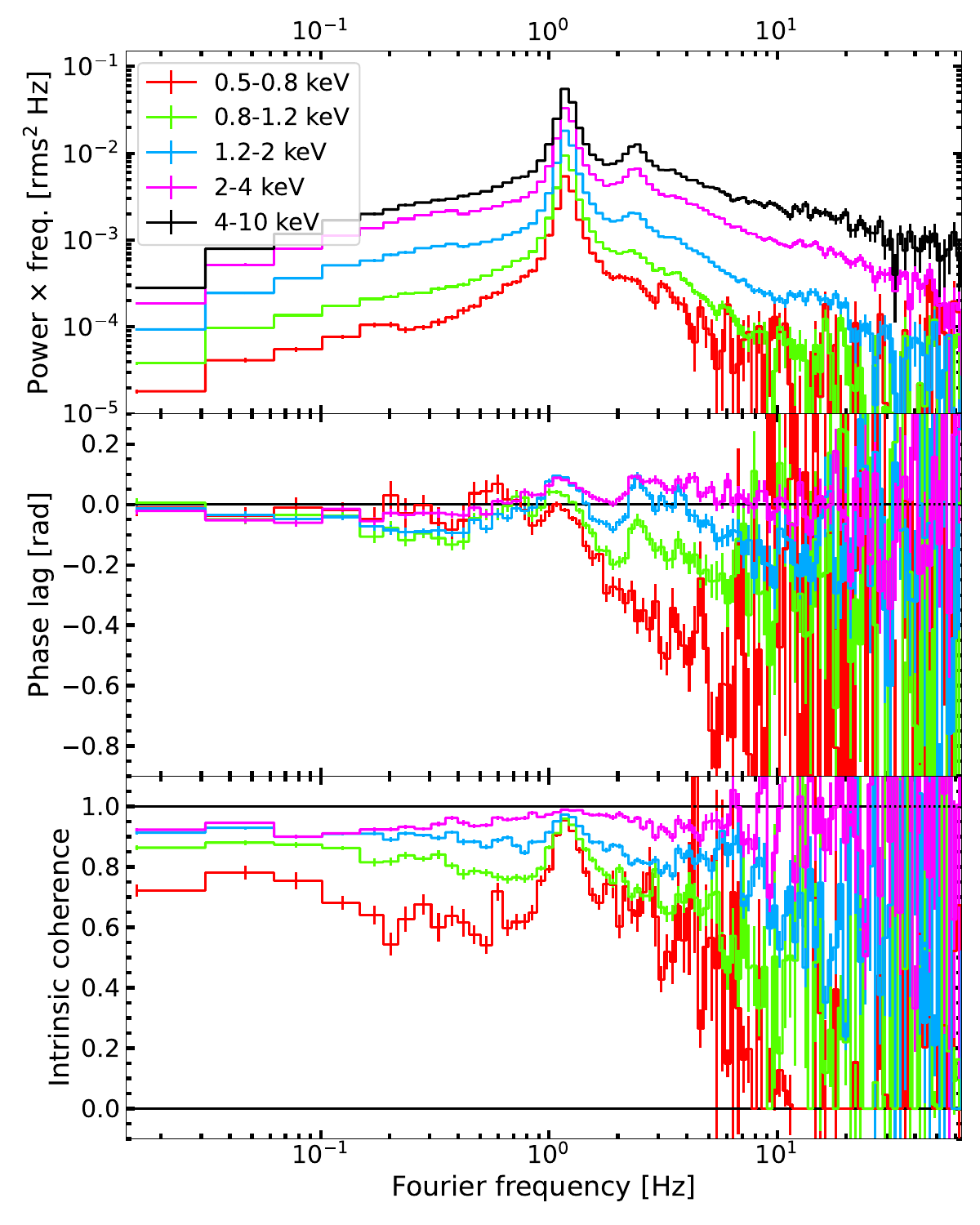}
    \caption{Power spectra and phase lag and coherence vs frequency spectra for different energy bands, using \nicer{} data with a QPO frequency of around 1.2 Hz for \source. For the phase lags and coherence, we use the 4-10 keV band as the harder band and the colour refers to the softer band. Above the QPO frequency, there are large soft lags when comparing a disc-dominated band with the powerlaw-dominated 4-10 keV band. The large change in fractional rms with energy is due to the soft photon contribution of the accretion disc, which varies much less than the corona, suppressing the amplitude of variability.}
    \label{fig:pslagcoh_nicer1.2}
\end{figure}

\begin{figure}
    \centering
    \includegraphics[width=\linewidth]{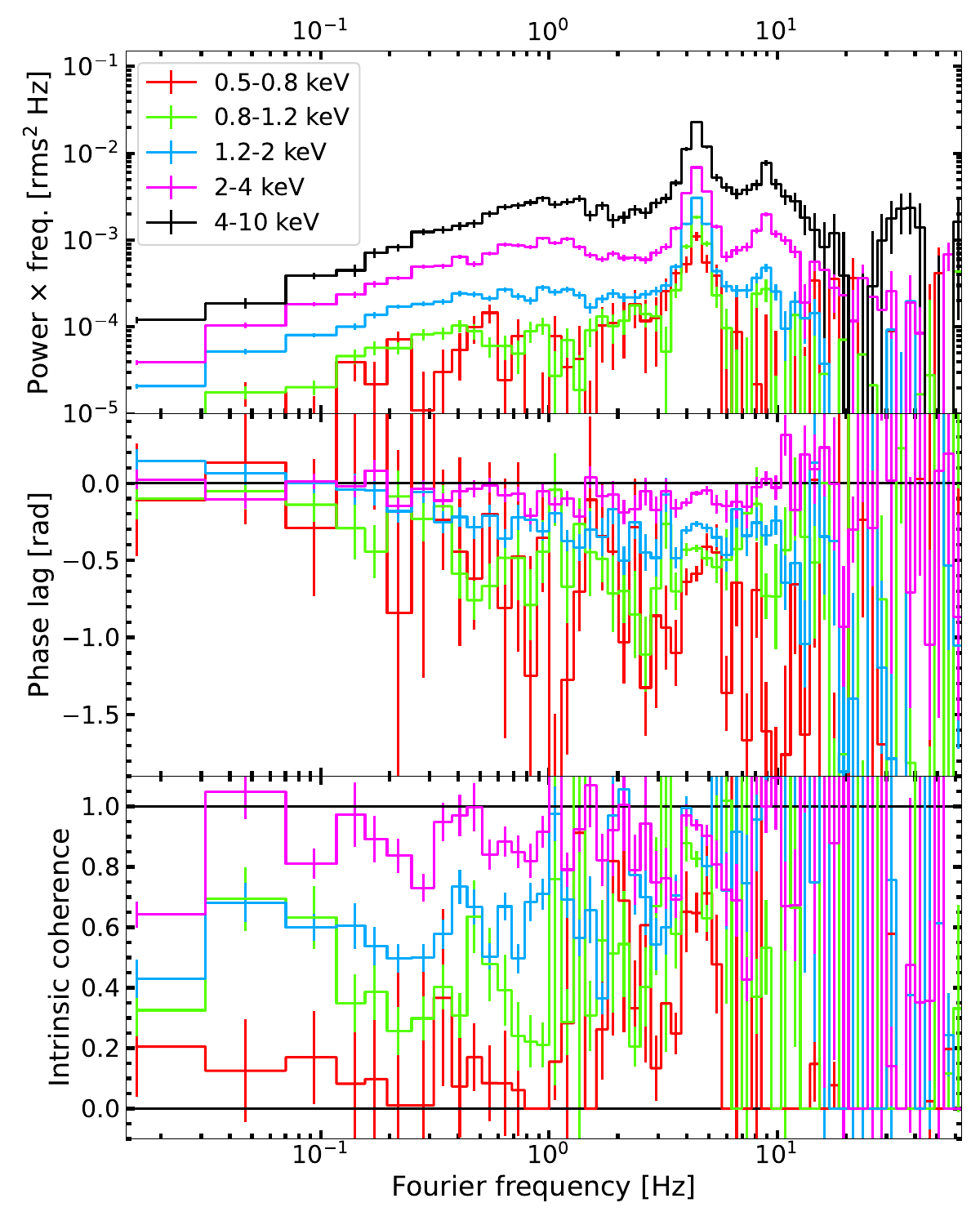}
    \caption{Power spectra and phase lag and coherence vs frequency spectra for different energy bands, using \nicer{} data with a QPO frequency of around 4.5 Hz, shown in a similar fashion to Fig. \ref{fig:pslagcoh_nicer1.2}.}
    \label{fig:pslagcoh_nicer4.5}
\end{figure}

\begin{figure}
    \centering
    \includegraphics[width=\linewidth]{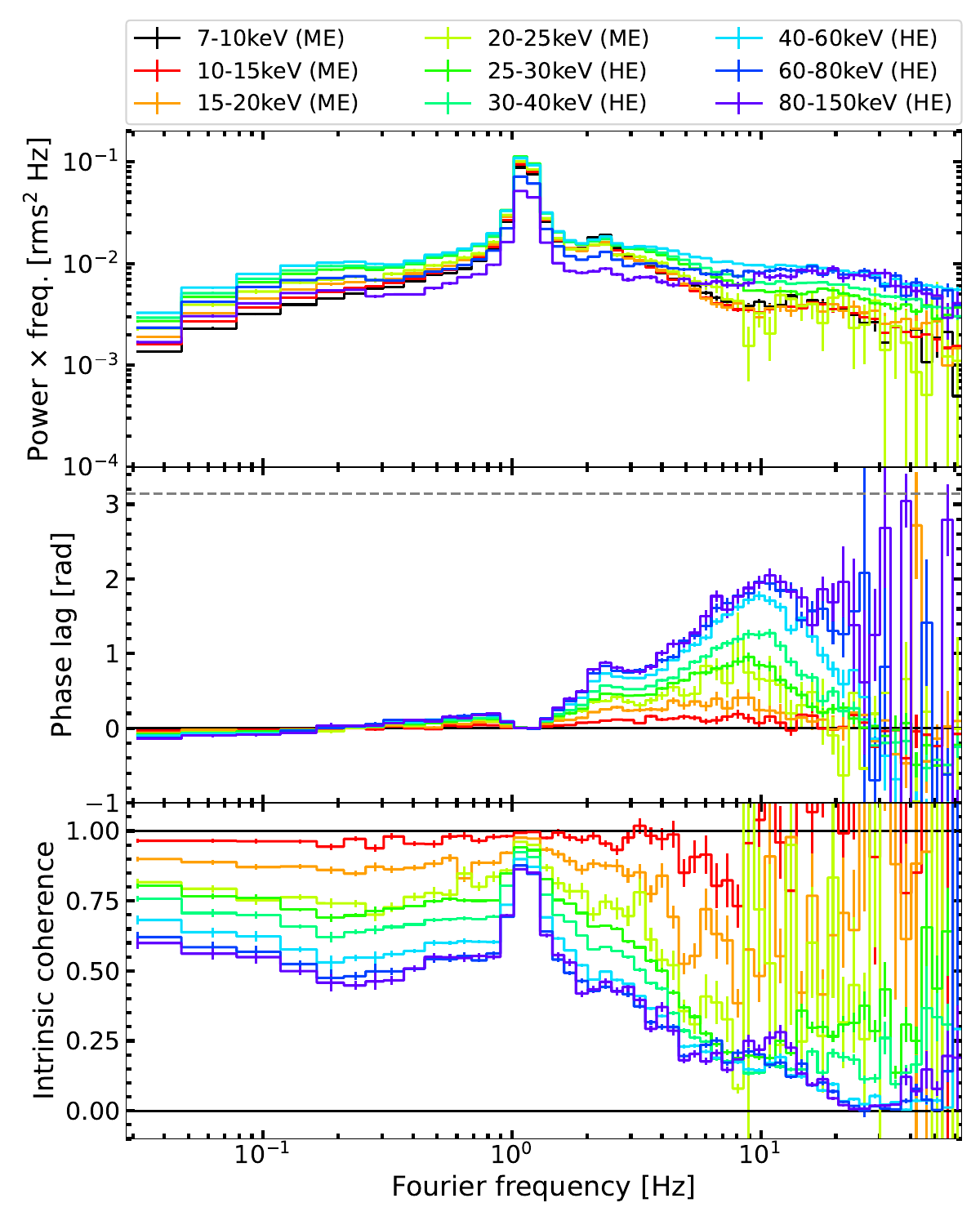}
    \caption{Power spectra and phase lag- and coherence vs frequency spectra for different energy bands, using \hxmt{} data with a QPO frequency of around 1.1 Hz for \source. The reduced fractional rms for the higher energies is at least partly artificial and due to the increased background contribution in those energy bands.}
    \label{fig:pstimelagcoh1.1}
\end{figure}

\begin{figure}
    \centering
    \includegraphics[width=\linewidth]{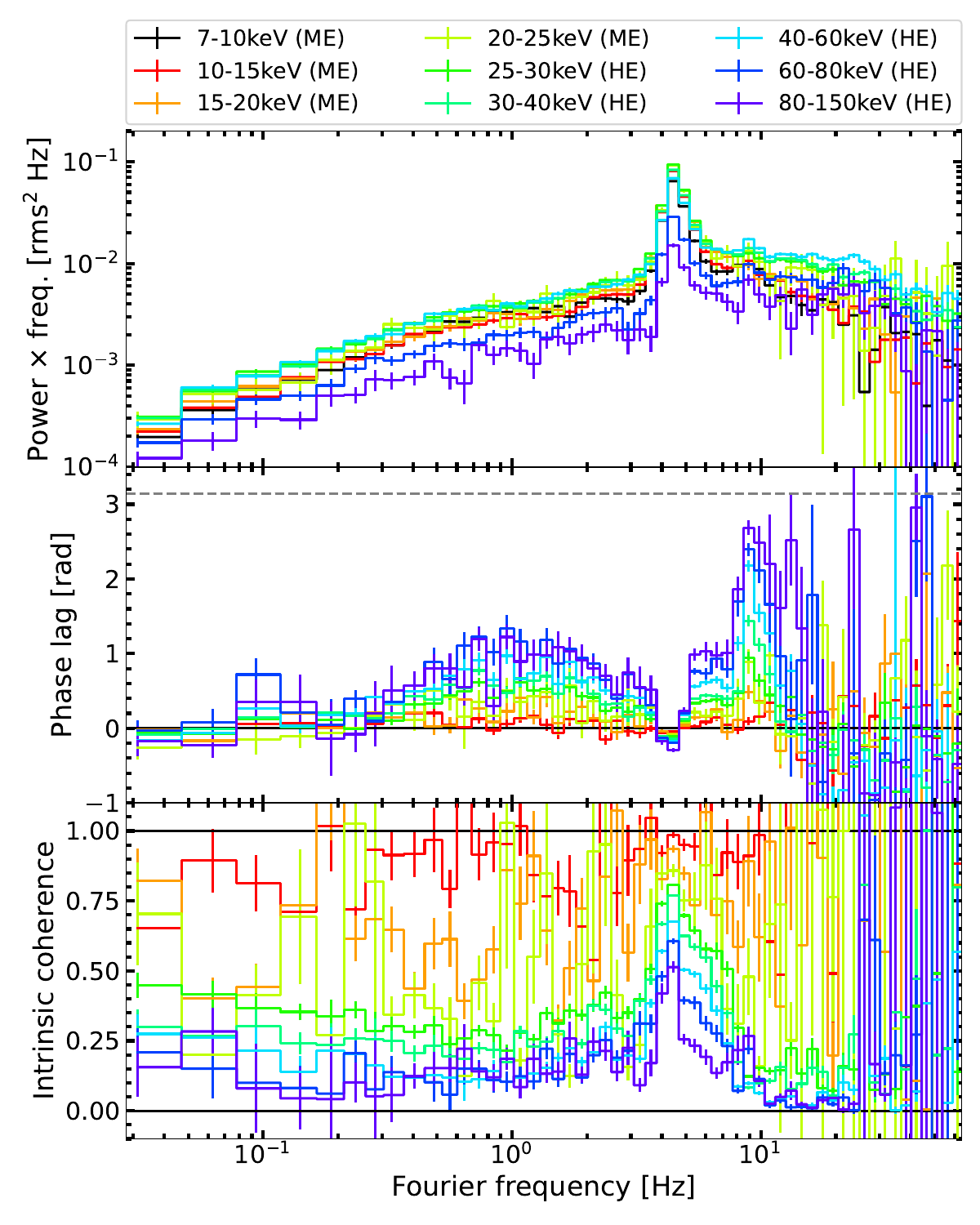}
    \caption{Power spectra and phase lag- and coherence vs frequency spectra for different energy bands, using \hxmt{} data with a QPO frequency of around 4.5 Hz, shown in the same way as in Fig. \ref{fig:pstimelagcoh1.1}. The reduced fractional rms for the hardest energy bands is at least partly artificial and due to the increased background contribution.}
    \label{fig:pstimelagcoh4.3}
\end{figure}

\begin{figure}
    \centering
        \includegraphics[width=\linewidth]{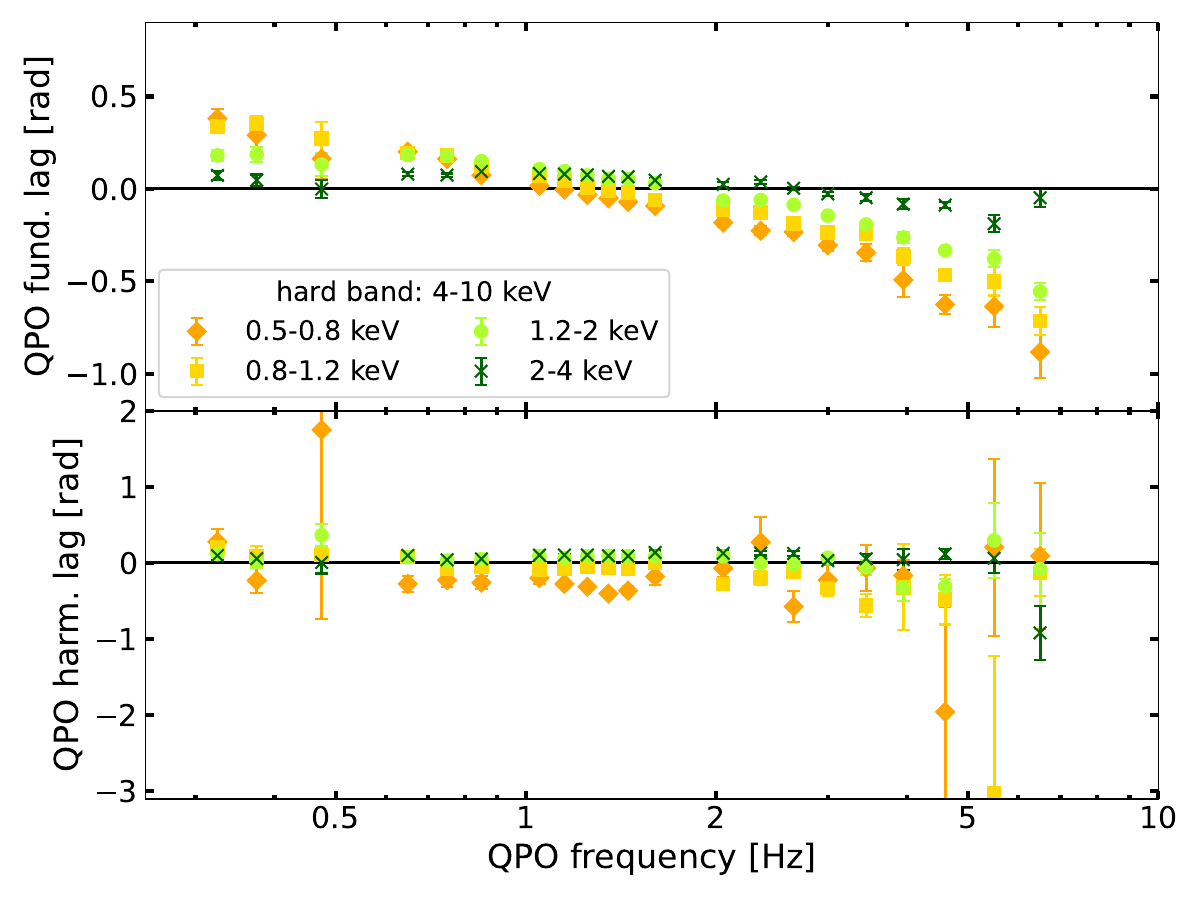}
    \caption{The upper panel shows the QPO fundamental phase lags binned on QPO frequency, while the lower panel shows the phase lags at the harmonic frequency for the \nicer{} observations of \source. Note that we use the convention here that positive lags are hard lags (hard photons lag soft photons), so technically, we are shifting the reference band energies. Compared to Fig. \ref{fig:QPO_lags_J1727}, we see that at low energies, the soft lags are much larger and decrease with QPO frequency. The harmonic lags are difficult to measure reliably at higher frequencies, but do not show a similar trend. }
    \label{fig:qpolag_nicer}
\end{figure}

\begin{figure*}
    \centering
    \includegraphics[width=175mm]{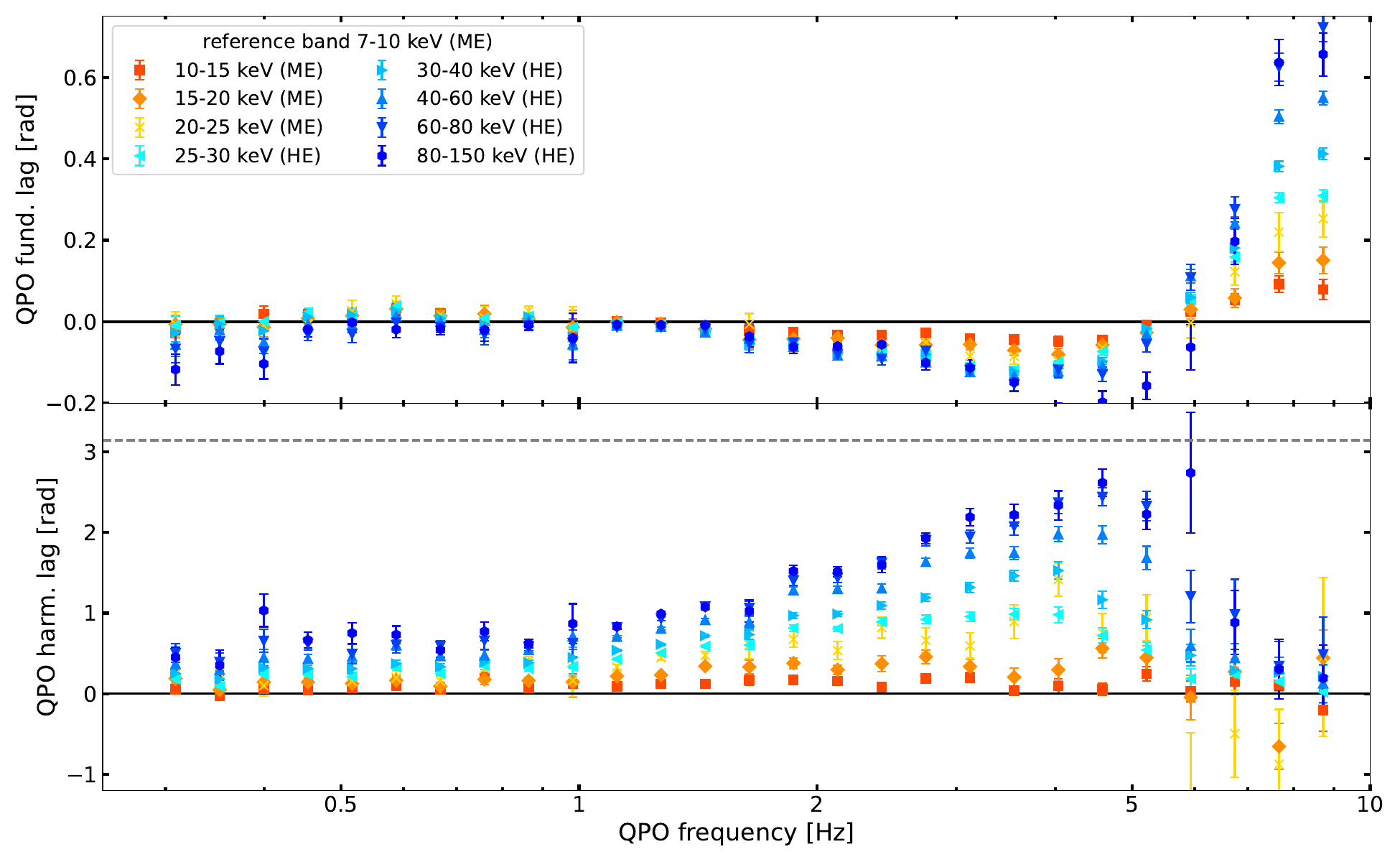}
    \caption{The phase lags at the QPO fundamental and harmonic frequencies for Swift J1727.8-1613 using \hxmt{} data, for different QPO frequencies and energy bands. The reference band is always the 7-10 keV band of the ME instrument. The orange to yellow colours are energy bands of the ME instrument, while the blue colours correspond to the HE instrument. The QPO fundamental lags are generally small, $\lesssim0.1$ rad, and progress from positive (hard) lags to negative between 1.5 and 5 Hz, becoming positive again at $>6$ Hz. At higher frequencies, the lags are larger for harder energy bands. The lower panel shows the harmonic lags, which look very different. The lags are always hard and increase with frequency and energy up to a QPO frequency of $\sim5$ Hz, while the lags decrease to small values around zero at higher QPO frequencies above $\sim$7 Hz.}
    \label{fig:QPO_lags_J1727}
\end{figure*}

\begin{figure}
    \centering
    \includegraphics[width=\linewidth]{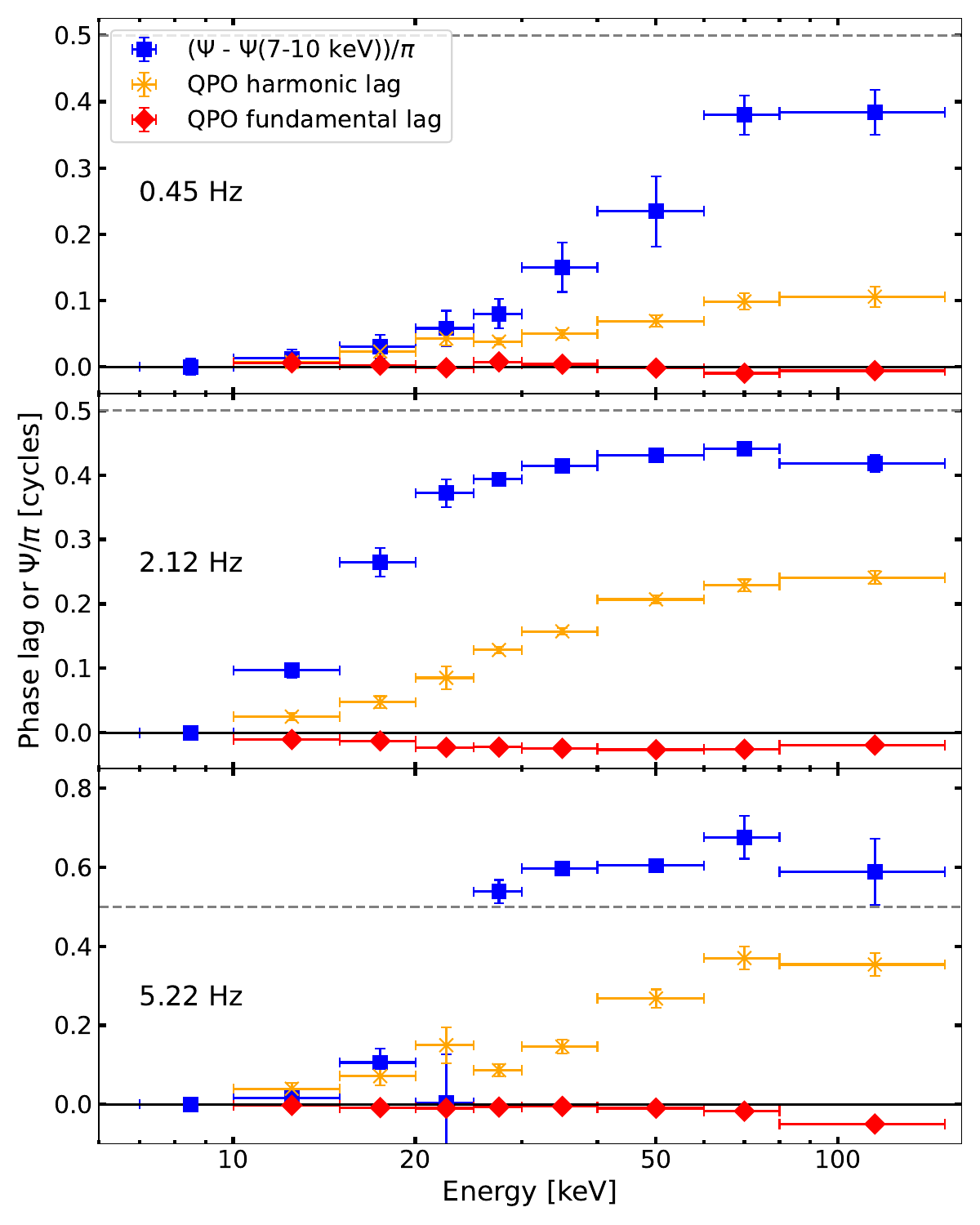}
    \caption{The panels show the energy dependence of phase difference $\Psi$ and the fundamental and harmonic phase lags for three different QPO frequencies using \hxmt{} data for \source. The reference band for the phase lags is 7-10 keV and they are measured only at the QPO fundamental or harmonic frequency, so in a single frequency bin. The value of $\Psi$ measured for the 7-10 keV band is subtracted from the measurements at higher energies to enable comparison between $\Psi$ and the phase lags at the harmonic. All values are given in units of cycles at the harmonic frequency, so the actual fundamental phase lags are a factor of 2 smaller than shown here. The dashed line at 0.5 shows the expected value of $\Psi$ above the pivot energy for a simple pivoting component (see Section \ref{sec:discussion_waveform}).}
    \label{fig:Psi_phlag_eng}
\end{figure}

\subsection{QPO lags}
\label{subsec:qpolag}

While the bispectrum can be used to study the phase relations between different frequencies in a single light curve, the cross-spectrum reveals the relation between two light curves of different energies at the same Fourier frequency. For a fully coherent pair of light curves, a difference in the biphase between two frequencies (so also in $\Psi$) in two separate energy bands will be visible as an energy-dependence of phase lags at either (or both) of the fundamental and harmonic frequencies \citep{Nathan_2022}. To investigate whether the strong dependence on energy of the biphase and thus the QPO waveform can be explained by strong energy-dependent lags, we studied the cross-spectral properties of \source{} in this section.

In Figs. \ref{fig:pslagcoh_nicer1.2}, \ref{fig:pslagcoh_nicer4.5}, \ref{fig:pstimelagcoh1.1} and \ref{fig:pstimelagcoh4.3}, we show example power spectra, phase lag-frequency and coherence-frequency spectra in different energy bands for \nicer{} (Figs. \ref{fig:pslagcoh_nicer1.2} and \ref{fig:pslagcoh_nicer4.5}) and \hxmt{} data (Figs. \ref{fig:pstimelagcoh1.1} and \ref{fig:pstimelagcoh4.3}). In Figs. \ref{fig:pslagcoh_nicer1.2} and \ref{fig:pslagcoh_nicer4.5}, we show the phase lags and coherence for four softer bands below 4 keV and the 4-10 keV energy range as hard band. We see a strong QPO at 1.2 and 4.5 Hz, respectively, at all energies, with the harmonic more prominent in the harder energy bands. There are large soft lags at high frequencies and on almost all timescales, the coherence is decreased for the softest energy bands, but it clearly peaks at the QPO frequency.

With the ME and HE instruments of \hxmt{}, using 7-10 keV as the reference band and harder energy ranges up to 150 keV as the subject bands, we see important differences between Fig. \ref{fig:pstimelagcoh1.1} and Fig. \ref{fig:pstimelagcoh4.3} as the QPO frequency increases. For a lower QPO frequency of 1.1 Hz, the low-frequency lags are small and there are very large and energy-dependent high-frequency lags, visible as a broad feature ranging from $\sim1.5$ to $\sim20$ Hz. The coherence at non-QPO frequencies decreases with energy and shows a clear peak at the QPO frequency. For a higher QPO frequency of 4.5 Hz, the large hard lags are visible both below and above the QPO frequency. The coherence is much lower, but again peaks at the QPO frequency. In the \hxmt{} data between 7 and 150 keV, the lags at the QPO frequency are much smaller than the broadband-noise-dominated surrounding frequencies, resulting in a clear dip in the phase lag vs frequency spectrum. 

The lags at the QPO frequency may contain information about the mechanism causing the QPO. \citet{VandenEijnden_2017} showed with archival RXTE data that the sign of the QPO fundamental phase lags generally depends on the inclination at which we view the system, an effect that becomes apparent as the QPO frequency increases. \source{} showed a broad range of QPO frequencies during its 2023 outburst, allowing us to do a similar analysis using \nicer{} and \hxmt{} data, extending the probed energy range from 2-13 keV in \citet{VandenEijnden_2017} to 0.5-150 keV. 

In Fig. \ref{fig:qpolag_nicer}, we show how the phase lags of the QPO fundamental and harmonic change with QPO frequency using \nicer{} data, while Fig. \ref{fig:QPO_lags_J1727} contains those for the ME and HE instruments of \hxmt. To measure the lags for all snapshots or sub-observations with a similar QPO frequency, we used the light curves mentioned in Section \ref{sec:qpowaveform}, which have a segment length of eight times the QPO period, equal to the expected length of a coherent interval \citep{Ingram_2015}. We calculated the lag at the Fourier frequency bin centered on the QPO frequency by obtaining the cross-spectrum between two energy bands and averaging over it for all segments \citep{Uttley_2014review}. We followed the same procedure for the harmonic lags, using the cross-spectrum at the Fourier frequency centered on $2\rm{\nu}_{QPO}$.

Fig. \ref{fig:qpolag_nicer} shows the QPO lags for different \nicer{} soft bands with respect to the 4-10 keV band. At low QPO frequencies (<1 Hz), the fundamental lags in the upper panel of Fig. \ref{fig:qpolag_nicer} are hard and increase with decreasing energy. As the QPO frequency increases, the fundamental lags become negative and show a strong energy dependence, where the lowest energy bands have the largest soft lags. The negative lag amplitude can be up to almost 1 rad for QPO frequencies of 6-8 Hz and for a very soft 0.5-0.8 keV and hard 4-10 keV band. The fact that the softest energies exhibit much larger changes as the QPO frequency evolves may be indicative of an effect of the QPO mechanism on the disc, possibly through reprocessed coronal emission on the disc. For example, if the QPO arises due to precession of the corona, the coronal emission may peak at a different phase than the illumination of the disc \citep{Stevens_2016,You_2020}. The QPO fundamental lags between powerlaw-dominated energy could therefore be small, while the disc-powerlaw lags evolve much more as the coronal geometry changes. For the harmonic frequency, the signal is less strong, but still shows a clear energy-dependent lag up to a QPO frequency of $\sim$5 Hz. The energy bands below 1.2 keV, where there is a strong contribution of the accretion disc, show negative lags, while the energy bands above 1.2 keV generally show (small) positive lags. At higher frequencies, the large uncertainties make it difficult to draw conclusions.

In \source, the QPO fundamental lag behaviour is similar to the trend observed by \citet{VandenEijnden_2017} for high-inclination sources, but it depends strongly on the soft energy band that is used to calculate the lags. All energies show similar evolution, with hard lags at low QPO frequencies and softer lags with increasing QPO frequency. However, the amplitude of the lags is highly energy-dependent, with softer energies having larger amplitudes. 

Fig. \ref{fig:QPO_lags_J1727} shows the energy-dependence of the QPO lags in \hxmt{} data and the following properties stand out. First of all, the QPO fundamental lags between 7-10 keV and higher energy bands are small, at most a few tenths of a radian. Secondly, they show a pattern of starting out with $\sim$0 rad for low (<2 Hz) QPO frequencies, becoming soft between 2 and 6 Hz and then switching to hard lags again for even higher frequencies. Above 6 Hz, the hard lag amplitude depends strongly on energy, with the highest energies showing the largest lags.

The picture is very different for the lags at the harmonic frequency, visible in the lower panel of Fig. \ref{fig:QPO_lags_J1727}. The lags are always hard and increase with QPO frequency up to about 5 Hz, with the hardest energy bands showing phase lags of almost $\pi$ rad, which would imply the presence of an anti-correlation. At even higher QPO frequencies, the harmonic lags return to small (hard) values. It must be noted that we do not see a bifurcation between energy bands below and above $\sim$15 keV, like we see in the QPO waveform. The smooth energy-dependence of the phase lags shows that the different methods (biphase and cross-spectral lags) pick out different parts of the signal at the same Fourier frequency.

The energy-dependence of the QPO lags and phase difference $\Psi$ are presented in a clearer way in Fig. \ref{fig:Psi_phlag_eng}, which shows the QPO lags and $\Psi$ vs energy for three QPO frequencies. All quantities are shown in units of cycles at the harmonic frequency and we subtract the 7-10 keV measurement for $\Psi$ off the other energy bands to enable comparison between the values at different energies. For fully coherent signals, we would expect that any difference in the QPO lags between energy bands translates into a corresponding difference in $\Psi$. However, it is clear from Fig. \ref{fig:Psi_phlag_eng} that the measured lags at the fundamental and harmonic frequencies are not large enough to explain the difference in $\Psi$. There is a much larger change in $\Psi$ with energy than in the measured phase lags, implying that the measured signal at the harmonic frequency must consist of multiple components, e.g. broadband noise and QPOs.

Looking at power spectra and lag-frequency spectra of harder bands and the 7-10 keV reference band, as shown in Figs. \ref{fig:pstimelagcoh1.1} and \ref{fig:pstimelagcoh4.3}, we see that the harmonic is weak in the power spectrum and the lags seem to be dominated by a much broader feature of hard lags at frequencies above the QPO fundamental and ranging up to $\sim$10-20 Hz. Such a high frequency lag feature for hard X-ray bands was observed earlier in MAXI J1820+070 with \hxmt{} data by \citet{WangYanan_2020} and \citet{Kawamura_2023} and in \source{} by \citet{Jin_2025}, who fitted the power- and cross-spectra with a multi-Lorentzian model. Proposed explanations for the large lags are Comptonisation delays and power-law pivoting, but the lags remain challenging to reproduce quantitatively with a physical model. The strong presence of the high-frequency lag feature suggests that the phase lags at the harmonic frequency lags in Fig. \ref{fig:QPO_lags_J1727} are probably not a good measure of the energy dependence of the intrinsic QPO harmonic phase. The fundamental is very strong during most of the outburst of \source{} and the lags seem to be dominated by a feature as broad as the QPO in the lag-frequency spectrum, at least at QPO frequencies $\lesssim$5 Hz. The fundamental lags might therefore be more reliable and point to a small intrinsic lag of the QPO fundamental itself, which can be used to compare to models. Because the broadband noise covers a wide range of frequencies, it is also present at the QPO frequency and will thus influence the lags.

When we compare the \nicer{} QPO lags to the results from measuring the waveform, we find a similar result. The waveform is similar for all \nicer{} energy bands, at least up to 6 Hz (see Fig. \ref{fig:Psi_J1727NICER}). Following the reasoning presented in \citet{Nathan_2022}, the phase lags at the fundamental and harmonic frequencies should be consistent with a similar phase difference $\Psi$ at different energies. In the \hxmt{} data, the signal was found to be more complicated and the phase lags cannot explain the energy-dependence of $\Psi$ (e.g. see Fig. \ref{fig:Psi_phlag_eng}). For simple coherent pairs of light curves with a similar value of $\Psi$ at different energies, we expect the lags at the QPO fundamental and harmonic to either not vary or to see the harmonic lags evolving in the same direction as the fundamental lags, with twice the fundamental phase lag amplitude. In the upper panel of Fig. \ref{fig:qpolag_nicer}, the QPO fundamental lags become softer as the QPO frequency increases. For such a change, the harmonic lags should evolve in the same direction and twice the amplitude to keep the waveform similar at all energies. We do not see such a pattern, although the lower rms at high frequencies in \nicer{} data, combined with the reduced number of detectors, do not allow measuring the harmonic lags reliably, especially above a few Hz. A detailed comparison between the cross-spectral phase lags and the biphase such as shown in Fig. \ref{fig:Psi_phlag_eng} is therefore not possible for \nicer{} data. Still, the large and energy-dependent changes in the phase lags, while the phase difference $\Psi$ does not depend on energy, show that the bispectrum and cross-spectrum probe different variability components in both \nicer{} and \hxmt{} data. Consistent with such a complicated signal, the cross-spectral coherence between different energies at the harmonic frequency is clearly lower than unity in Figs. \ref{fig:pslagcoh_nicer1.2}-\ref{fig:pstimelagcoh4.3}.

From these considerations, we conclude that the cross- and bispectrum extract separate variability components at the same Fourier frequency. The phase lags measured with the cross-spectrum at the harmonic frequency are likely dominated by the broadband noise lags that are visible over a broader frequency range. The biphase, however, extracts the phase relation between the QPO fundamental and harmonic and is therefore an independent probe of the QPO mechanism \citep{Maccarone_2013}. Using the cross-spectrum and the bispectrum, we can separate distinct components of the complex variability produced by the system.

\subsection{Comparison to \sourcem{}}
\label{sec:discussion_1535}

\begin{figure}
    \centering
    \includegraphics[width=\linewidth]{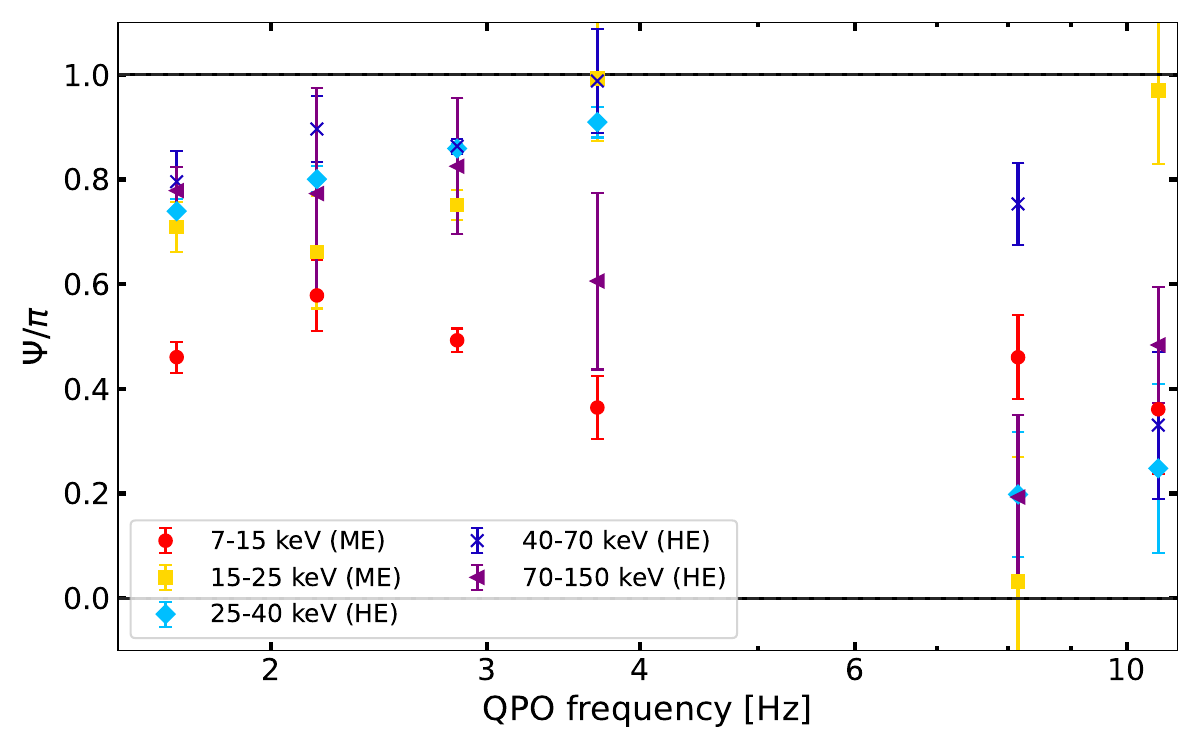}
    \caption{The phase difference $\Psi$ for \sourcem{} was calculated from \hxmt{} data, using the biphase at the QPO fundamental and harmonic frequencies for different QPO frequencies and energy bands. The general trends are similar to those in Fig. \ref{fig:Psi_J1727}, although the values of $\Psi$ are slightly higher for \sourcem{} compared to \source. Below 15 keV, $\Psi$ decreases with frequency, while above 15 keV, there is a clear increase. There were no observations with QPO frequencies between 4 and 8 Hz.}
    \label{fig:Psi_J1535}
\end{figure}

\begin{figure}
    \centering
    \includegraphics[width=\linewidth]{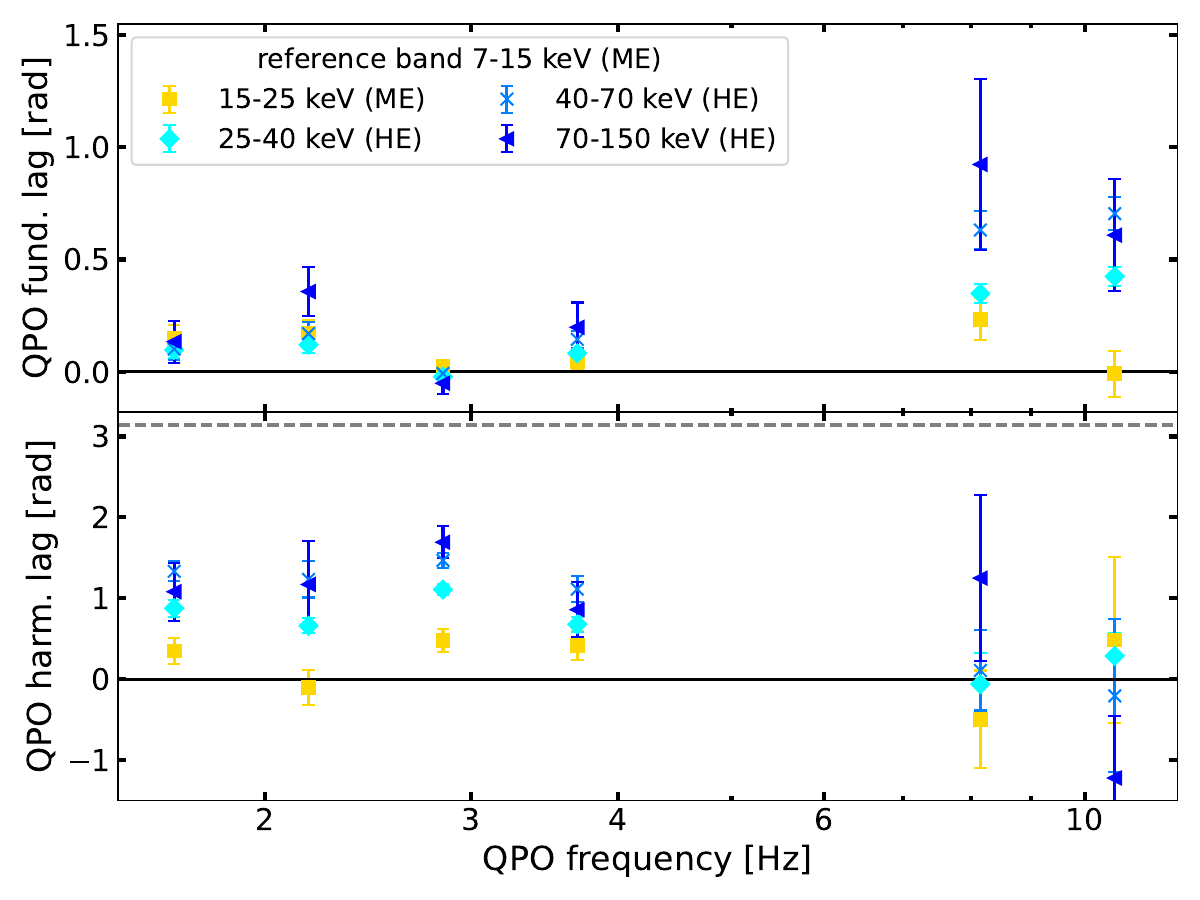}
    \caption{The phase lags at the QPO fundamental and harmonic frequencies for \sourcem{} using \hxmt{} data, for different QPO frequencies and energy bands. The reference band is always the 7-15 keV band of the ME instrument. Since \sourcem{} has much sparser coverage, we use broader energy bands than in Fig. \ref{fig:QPO_lags_J1727}, but the general lag pattern is consistent with what we measure in \source.}
    \label{fig:QPO_lags_J1535}
\end{figure}

\begin{figure}
    \centering
    \includegraphics[width=\linewidth]{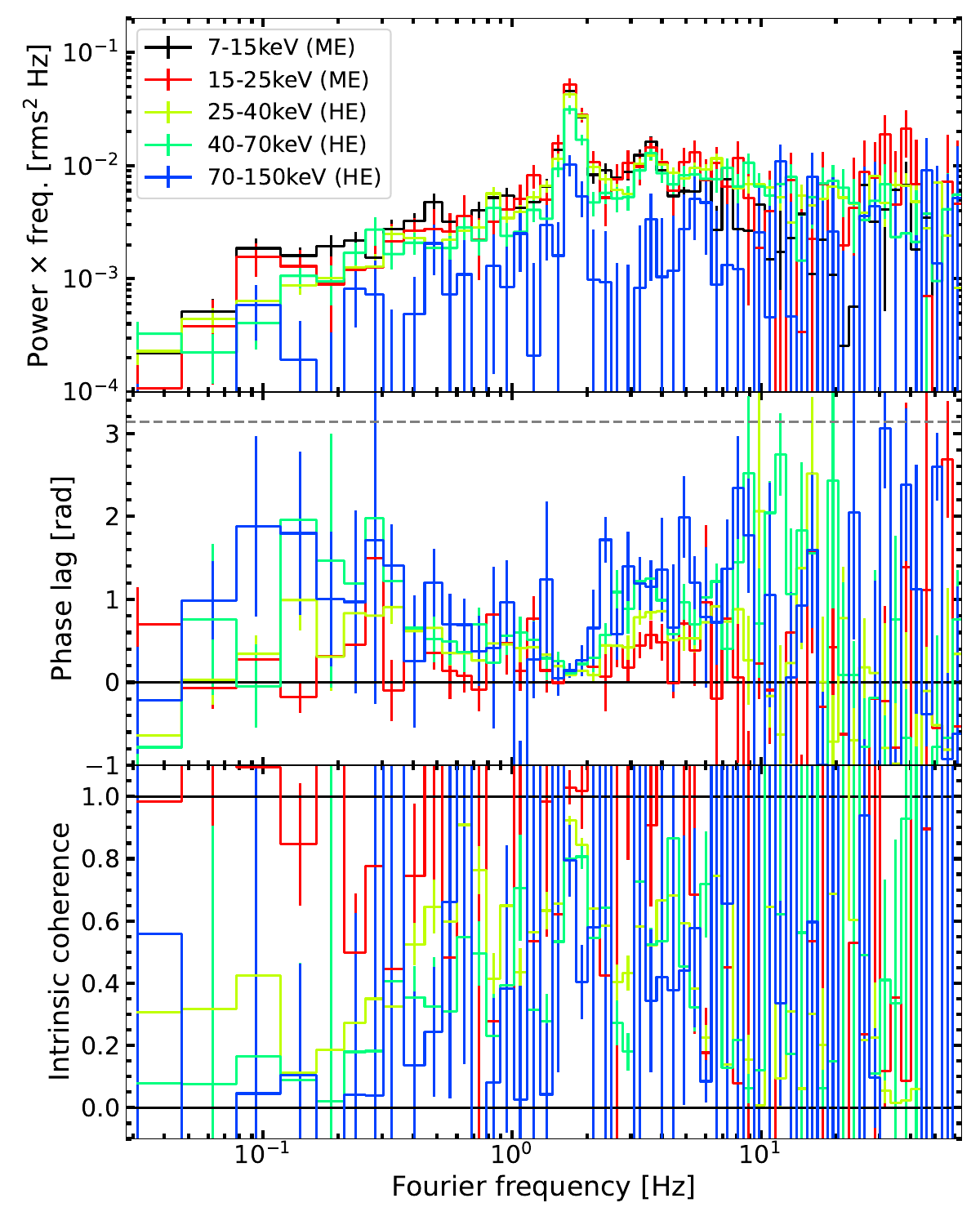}
    \caption{The three panels show the power spectra, phase lag and coherence vs. frequency spectra for different energy bands of \hxmt{} observations of for \sourcem{} with a QPO frequency of $\sim1.7$Hz. The general pattern in all three properties is very similar to what we observe for \source{} at high QPO frequencies, as shown for example in Fig. \ref{fig:pstimelagcoh4.3}. The power spectra change little with energy and show a strong QPO. The phase lag vs. frequency spectra show an energy-dependent continuum of broadband noise lags, while the lags at the QPO frequency are close to zero. The broadband coherence is hard to constrain, but clearly peaks at the QPO frequency and is decreased at other frequencies for the harder energy bands.} 
    \label{fig:pstimelag1535}
\end{figure}

As is clear from Figs. \ref{fig:Psi_J1535} and \ref{fig:QPO_lags_J1535}, the QPO waveform and phase lags as a function of energy behave similarly in \source{} and \sourcem. We used all available \hxmt{} data for the HIMS of \sourcem{} and applied the same analysis as laid out in the previous paragraphs. The smaller number of observations leads to noisier measurements, but we can still discern similarities between the sources. Although the measured phase difference $\Psi$ and phase lags are not quite the same in both sources, the pattern that emerges as the QPO frequency increases is remarkably similar. Lower energies ($\lesssim$15 keV) decrease from $\Psi$/$\pi\sim$0.6 to $\sim$0.4 with QPO frequency, while higher energy bands show an increase in $\Psi$ with QPO frequency. In Fig. \ref{fig:pstimelag1535}, we show power spectra, phase lag and intrinsic coherence vs frequency spectra for different energy bands. For the two lower panels, the reference band is 7-15 keV. The phase lag vs. frequency spectra show broad structures with large amplitudes for high energies, but the lags are close to zero at the QPO fundamental frequency, very similar to what we observe in \source. The coherence at the QPO fundamental frequency is also high in both \source{} and \sourcem. Both sources transitioned through the hard state quickly \citep{Nakahira_2018,Ingram_2024}, show strong QPOs and are probably viewed at medium-high inclination angles of 57-67$\degree$ for \sourcem{} and 30-50 $\degree$ for \source{} \citep{Miller_2018,Xu_2018,Svoboda_2024,Peng_2024}. 

Apart from \sourcem, \citet{Ingram_2024} argue that the soft lag behaviour of \source{} agrees with BHXRBs MAXI~J1803$-$298 and EXO~1846$-$031 as reported by \citet{Wang_2022_revmachine}. The similarities between these sources indicate that the complex spectral-timing behaviour observed in unprecedented detail in \source{} is not unique and may at least be present in a subset of BHXRBs. We attempted to measure the QPO waveforms and lags in MAXI~J1803$-$298 and EXO~1846$-$031 as well, but their lower brightness and smaller number of \hxmt{} observations prevented us from drawing conclusions from those measurements. 

\section{Discussion}
\label{sec:discussion}

\subsection{Energy-dependent waveform}
\label{sec:discussion_waveform}

We propose that the energy-dependence of the QPO waveform presented in Figs. \ref{fig:Psi_J1727} and \ref{fig:Psi_J1727NICER} can be explained by power-law pivoting in the following way. It has been observed in the past that the sign of the lags alternates for different harmonics \citep{Cui_1999,Cui_2000,Casella_2004}, which is the case for \source{} as well. The fundamental lags are mostly close to 0 or negative, whereas the harmonic lags are always positive, at least at the coronal power-law dominated energies shown in Fig. \ref{fig:QPO_lags_J1727}. The alternating sign of the lags has inspired the idea that spectral pivoting is a key property of QPOs \citep{Shaposhnikov_2012,Misra_2013}. The authors of both papers show how a QPO signal in two independent parameters can lead to complex lag patterns and \citet{Misra_2013} further show an example of what the biphase could look like for different energies. Their Fig. 3 does not correspond to the sharp change we observe in \source, but the framework might be able to reproduce our findings. The pivot in power-law index $\Gamma$ was first confirmed with QPO-phase-resolved spectroscopy by \citet{Ingram_2015} for GRS 1915+105 and later for other sources \citep{Ingram_2016,Stevens_2016}. However, from these QPO-phase-resolved analyses, it is unclear what the pivot energy is.

\begin{figure}
    \centering
    \includegraphics[width=\linewidth]{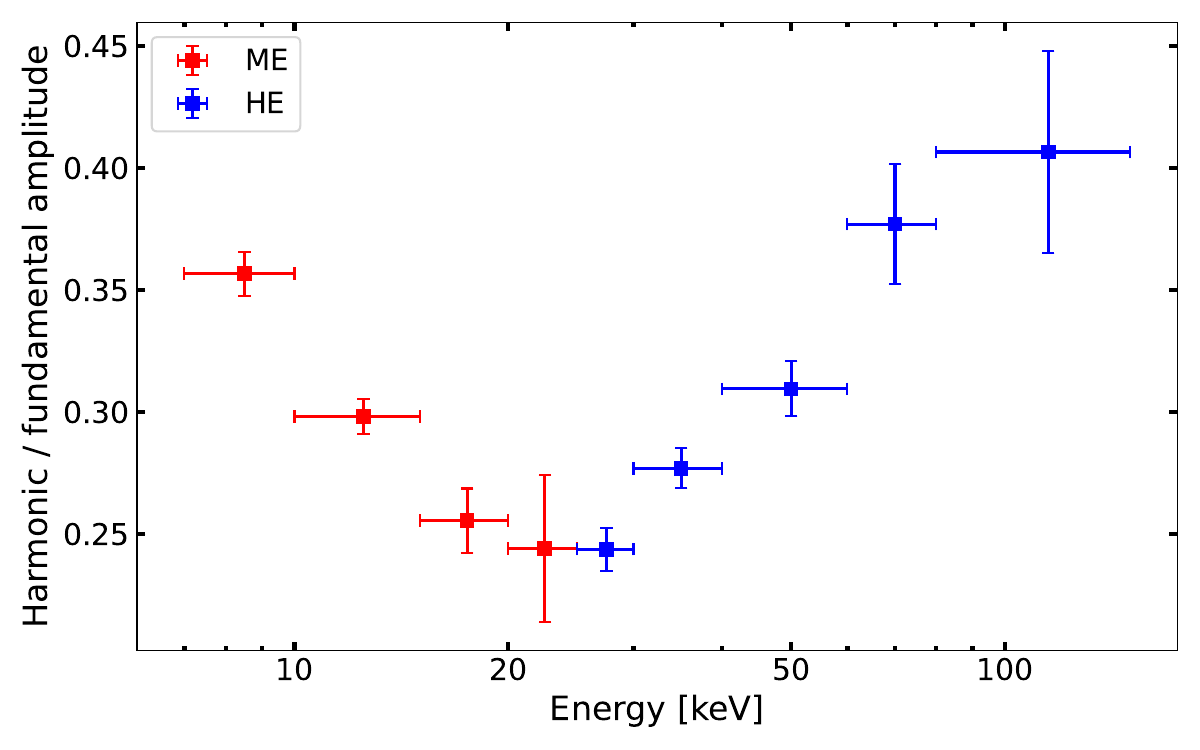}
    \caption{The ratio of the harmonic and the fundamental amplitude obtained by fitting Lorentzians to the power spectra of different energy bands for all \hxmt{} data with a QPO frequency around 2.4 Hz for \source. There is a clear minimum at $\sim$20 keV, which is expected around the pivot energy.}
    \label{fig:ampratio_energy}
\end{figure}

\begin{figure}
    \centering
    \includegraphics[width=\linewidth]{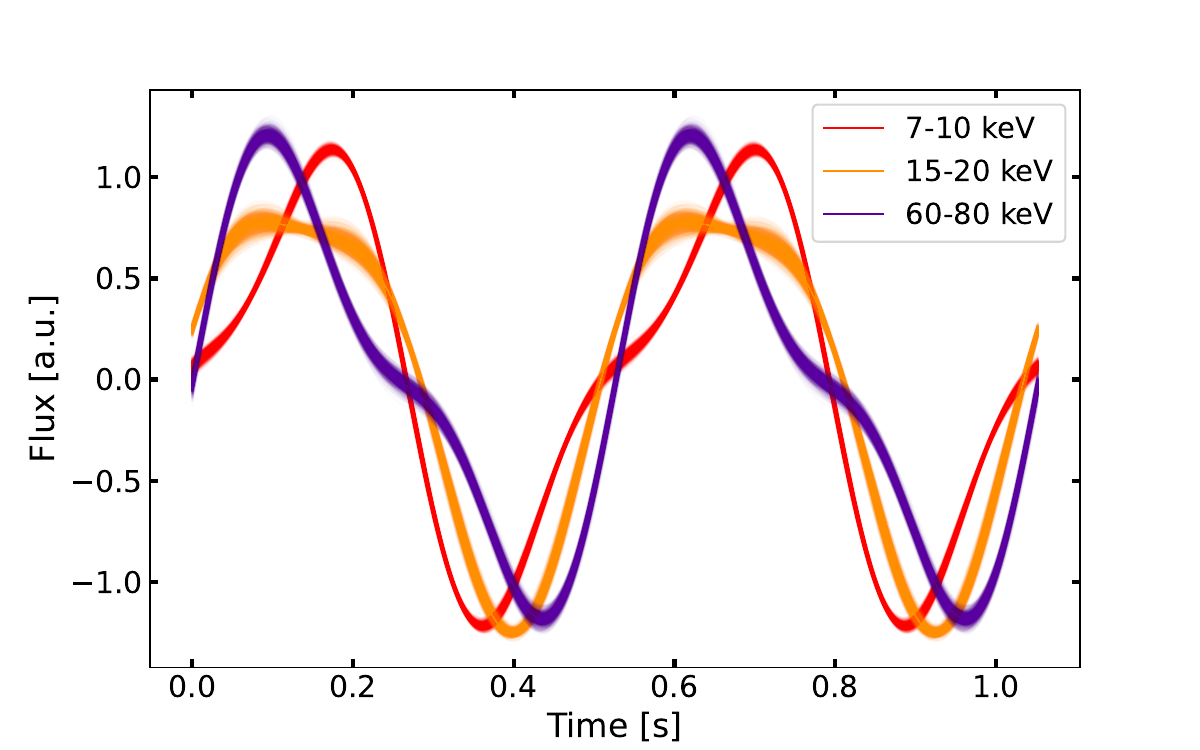}
    \caption{Examples of the inferred waveform for three different energy bands, using \hxmt{} data with a QPO frequency around 2.4 Hz for \source. The errors are represented by the distribution of the 1000 thin lines for each energy band, which were created by randomly drawing 1000 values from the amplitude ratio and phase difference parameter distributions, assuming normally distributed errors. Each line represents a waveform with randomly drawn parameters. There is a clear and significant dependence of the waveform on the energy band used. For illustrative purposes, we assume the fundamental lags are zero here, which is close to the values measured and shown in Fig. \ref{fig:QPO_lags_J1727}.}
    \label{fig:waveform_example}
\end{figure}

The symmetry of the increase ($>$20 keV) and decrease ($<$15 keV) of $\Psi$ with QPO frequency, especially between $\sim$1.4 and 4 Hz, suggests that a property of the QPO waveform has an opposite sign below and above 15 keV. We propose here that what we may be seeing is a spectral component with the spectral shape of a power-law, pivoting at the harmonic frequency with a pivot energy of around 15-20 keV. Such a pivot will show less variability around the pivot energy and indeed we find that the strength of the harmonic relative to the fundamental has a minimum around those energies. In Fig. \ref{fig:ampratio_energy}, we show the ratio of the harmonic and the fundamental amplitudes for different energy bands in \source. We fitted two narrow Lorentzians to the fundamental and harmonic and tied their centroid frequencies such that the harmonic has twice the frequency of the fundamental. We also fitted three broad Lorentzians to cover the broadband noise. For all Lorentzians, we tied the centres and widths between different energy bands and let the amplitudes vary. It is clear from Fig. \ref{fig:ampratio_energy} that the strength of the harmonic relative to the fundamental has a minimum around 20 keV, which is where we expect the pivot point based on our energy-dependent waveform measurements. We do note that the broadband noise shape depends on energy and as such the power spectra of some energy bands require only two broad Lorentzians, which could artificially increase the strength of the harmonic. In Fig. \ref{fig:waveform_example}, we show example waveforms of three different energies for \hxmt{} data with a QPO frequency around 1.9 Hz. The errors are represented by the distribution of the narrow lines, which are determined by randomly drawing 1000 values from the parameter distribution of the fitted values of $\Psi$ and the amplitude ratio, assuming normally distributed errors. Fig. \ref{fig:waveform_example} clearly shows that the measured QPO waveform depends on energy.

By measuring $\Psi$ and the strength of the harmonic at different QPO frequencies and energies, we conclude that the pivot energy does not stay constant. At low frequencies, the pivot energy seems to be higher ($\sim$40 keV), which would explain the difference in $\Psi$ above and below 40 keV for QPO frequencies $\lesssim$0.5 Hz (see Fig. \ref{fig:Psi_J1727}). The energy-dependence of $\Psi$ is clearly visible in Fig. \ref{fig:Psi_phlag_eng}, which shows $\Psi$ and QPO lags vs energy for three different QPO frequencies. In the upper panel, for a QPO frequency of 0.45 Hz, the pivot energy seems to be around 40 keV, while observations with higher QPO frequencies are more consistent with pivot energies of about 20 keV. It is important to note that energy-dependence of the waveform is more complicated than a simple pivoting model can explain, as there is a gradual increase in $\Psi$ below the inferred pivot energy, while above the pivot energy, $\Psi$ is constant. If the biphase is a good probe of the harmonic signal, we infer there are extra delays in the harmonic signal below the pivot energy, which require a more complex model than the simple framework discussed here. Given the expected changes in the properties of the accretion flow as the QPO frequency evolves, it is not surprising that the pivot energy changes as well, and it may be an extra parameter that can be used to test models. Interpreting the energy-dependent waveform as being due to a pivoting power-law ties in with the results presented by \citet{Axelsson_2014}, who found that the strength of the harmonic decreases with energy up to 20 keV in XTE~J1550-564 using \rxte{} data, where it was not possible to probe the strength of harmonic at higher energies. Our results suggest that the harmonic amplitude may increase again at higher energies in XTE~J1550-564 as well.

A pivoting power-law at the harmonic frequency could lead to the waveform trends we observe in the following way. If we assume the QPO is due to a precessing corona, the main mechanisms for flux variations could (a) be the change in the solid angle of the corona as seen by the observer and (b) beaming effects as the blueshifted part of the corona becomes less or more visible during a precession cycle \citep{Ingram_2019review}. If we imagine that both of these effects have two maxima per QPO cycle, with one being much larger than the other and causing the QPO fundamental to be strongest, we naturally obtain a harmonic component that is linked to the QPO fundamental. If harmonic components a and b are not in phase with each other, they combine to make a single harmonic following the trigonometric identity

\begin{equation}
\label{eq:harm_waveform1}
    {A_h}\sin{(\theta+\phi_h)}={A_a}\sin{(\theta+\phi_a)}+{A_b}\sin{(\theta+\phi_b)},
\end{equation}
with ${A_h}$, ${A_a}$ and ${A_b}$ the amplitudes of the full harmonic, the solid angle component and the beaming component, respectively, while $\phi_h$, $\phi_a$ and $\phi_b$ are the associated total and component phases and $\phi_a \neq \phi_b$ to make sure the components are not in phase. If we assume that these effects simply add up to create the harmonic component of the waveform, we also find
\begin{equation}
    \tan{\phi_h} = \frac{{A_a}\sin{\phi_a}+{A_b}\sin{\phi_b}}{{A_a}\cos{\phi_a}+{A_b}\cos{\phi_b}}.
\end{equation}

One of these components may also be associated with a spectral pivot, for example due to changes in the ratio between the seed photon and heating luminosity as the coronal geometry as seen from the seed-emitting region (e.g. the inner disc) varies. If such a pivoting component contributes significantly to the waveform, the phase of the harmonic signal will depend on whether the energy band is lower or higher than the pivot energy. Below the pivot energy, the signal could be described by equation \ref{eq:harm_waveform1}, while above the pivot energy, the equation would be

\begin{equation}
\label{eq:harm_waveform2}
    {A_{h}}\sin{(\theta+\phi_h)}={A_{a}}\sin{(\theta+\phi_a)}+{A_{b}}\sin{(\theta+\phi_b+\pi)}.
\end{equation}

As the ratio of amplitude ratio of ${A_a}$ and ${A_b}$ changes with QPO frequency, which is probably related to longer term geometric changes in the coronal size or radius, the resultant $\phi_h$ also changes. A simple simulation shows that this scenario succeeds in producing the mirrored waveform behaviour, if the relative amplitudes of the solid angle and beaming components change strongly with QPO frequency. In fact, ${A_a}$ and ${A_b}$ do not need to be related to these two physical mechanisms, any pair of mechanisms that are not varying in phase, of which one has a pivoting component, will suffice.

\begin{figure}
    \centering
    \includegraphics[width=\linewidth]{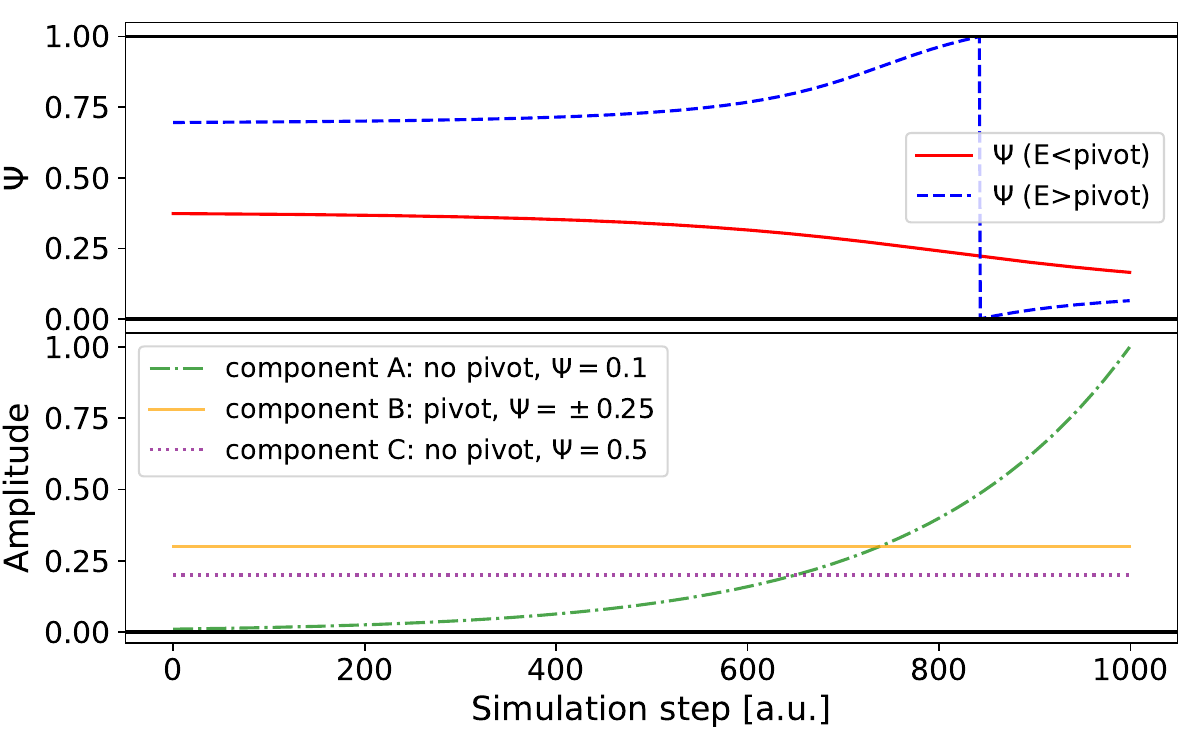}
    \caption{The two panels show how $\Psi$ changes for different amplitudes of three components A, B and C, which together create the QPO signal at the harmonic frequency. In the upper panel, the solid red line and the dashed blue line represent $\Psi$ below and above the pivot energy, respectively, and they follow the two main trends in Fig. \ref{fig:Psi_J1727}, for energy bands below and above the pivot energy. The lower panel shows the amplitudes and $\Psi$ values for the three distinct components.}
    \label{fig:Psi_sim}
\end{figure}

To reproduce the data, however, another modification is needed. To obtain the trends visible in Fig. \ref{fig:Psi_sim}, we need a third component that is in phase for all energies. We require such a component C in order to recreate the measured difference in $\Psi$ between energy bands below and above the pivot, which is smaller than 0.5 cycles at low QPO frequencies (see e.g. upper panels of Fig. \ref{fig:Psi_phlag_eng}). At the highest QPO frequencies, component A dominates to reproduce the observed waveform evolution. The above phenomenological interpretation is partly based on the discussion of \citet{De_Ruiter_2019}, who show it is possible to model a changing waveform by changing the relative strength of two underlying waveforms with fixed phase difference. Also, comparing our work to the detailed modeling of a precessing inner accretion flow presented in \citet{You_2018} and \citet{You_2020}, we find that complicated energy-dependent waveforms are to be expected, but reproducing our results in such a framework may require further exploration of the parameter space.

\subsection{Lags}
\label{sec:discussion_lags}

The broadband shape of the lag-frequency spectra shown in Figs. \ref{fig:pslagcoh_nicer1.2}, \ref{fig:pslagcoh_nicer4.5}, \ref{fig:pstimelagcoh1.1} and \ref{fig:pstimelagcoh4.3} develops strongly as the QPO frequency increases. At low QPO frequencies and high energies (Figs. \ref{fig:pstimelagcoh1.1} and \ref{fig:pstimelagcoh4.3}), we see modest hard or soft lags below the QPO frequency, while above the QPO frequency, there are large, energy-dependent hard lags. At higher QPO frequencies, strong hard lags emerge below $\rm{\nu_{QPO}}$, while the lags above the QPO stay similar or become smaller. The high frequency lags can go to very large values, close to $\pi$. In principle, powerlaw pivoting can produce lags with a log-linear energy dependence if there is a time delay between normalization and powerlaw index changes \citep{Koerding_2004}. \citet{Kawamura_2023} tried to model a similar high frequency lag feature in the \hxmt{} data of MAXI~J1820+070 and found that fitting the lags at the highest energies through powerlaw pivoting is non-trivial and requires improved physical prescriptions. Power-law pivoting can be induced by changes in the ratio of the number of seed photons available for Comptonisation and the heating of the corona, as was shown by \citet{Uttley_2025} in their model for lags in broadband noise variability. The size and sign of the lags depend strongly on the coronal geometry, as it determines the number and radius of origin of seed photons from the disc reaching the corona. 

The coronal geometry also plays a major role in the \texttt{vKompth} model for QPO lags \citep{Bellavita_2022}. In the model, the QPO lags are caused by Comptonisation delays and depend mainly on the coronal size and the feedback fraction, which determines how much emission from the corona is emitted towards the disc. The model was applied to several BHXRBs and can fit the energy-dependence of the QPO lags (e.g. \citealt{Zhang_2022,Rout_2023,Alabarta_2025}). When fitting the QPO lags in the presence of strong broadband noise, such as in \source, it is important to distinguish between lags that arise due to broadband noise that is also present at the QPO frequency, and intrinsic QPO lags that are due to the QPO mechanism. \citet{Mendez_2024} show that by fitting Lorentzians to power- and cross-spectra, it is possible to extract the intrinsic QPO lags, as was also demonstrated by \citet{Jin_2025} for the \hxmt{} data set of \source.

Following a different approach to differentiate between QPO and broadband noise contributions,\citet{Yu_2024}, after \citet{Ma_2021} and \citet{Zhou_2022}, assume that the broadband noise and QPO signals are coupled by convolution in the time domain. Also, the authors assume that the broadband noise lags at the QPO frequency are similar to those at sub-QPO frequencies. With those assumptions, it is possible to extract the intrinsic QPO lags by subtracting the mean (broadband noise) lag below the QPO frequency. Even if both assumptions are correct, which may be true as there is little evidence in favour or against the assumptions, their results imply the following. The fact that the measured QPO lag is small at almost all frequencies and the lower-frequency broadband noise lags change strongly implies that the QPO `knows' about the lags at longer timescales and the QPO lags at all energies have just the right size in the opposite (soft) direction to make the measured lags close to zero. From the high coherence at the QPO frequency (see Fig. \ref{fig:pstimelagcoh1.1}), it seems more likely that a single component, e.g. the QPO signal, dominates the lag measurement \citep{Vaughan_1997coherence}. The QPO fundamental lag is indeed close to zero in the energy range 7-150 keV for all but the highest QPO frequencies. Such a small lag for a broad range of energies may be compatible with the idea of a precessing corona, if the main high energy flux modulation is dominated by solid angle differences between precession phases and happens simultaneously across different energies, providing a constraint for the origin of the hardest X-rays in BHXRBs. At softer X-ray energies, in the \nicer{} band, the QPO fundamental lags depend more strongly on energy, as is visible in Fig. \ref{fig:qpolag_nicer}, which may be related to the accretion disc. A precessing corona leads to a varying illumination pattern of the accretion disc, which can cause lags \citep{Stevens_2016}. In any case, more detailed precession models which can make spectral-timing predictions are necessary to investigate whether precession can explain the observed lags. 

As was shown above, several models indicate that the geometry of the corona changes as the QPO frequency increases, and the details of such a geometric change may be key to understanding the complex lag behaviour below, at and above the QPO frequency (e.g. \citealt{Kara_2019,Mendez_2022}). On much shorter time-scales of seconds, \citet{Bollemeijer_2024} observed that in MAXI~J1820+070, the soft lags between disc- and corona-dominated energy bands show significant variation, implying that the corona is a dynamic structure with a geometry that varies on a broad range of time-scales. Changes in the lags on time-scales shorter than the light curve segment length will cause the coherence to decrease \citep{Nowak_1999}. We note that in Figs. \ref{fig:pstimelagcoh1.1} and \ref{fig:pstimelagcoh4.3}, the frequency-dependent intrinsic coherence at non-QPO frequencies decreases with energy. The observed low coherence indicates either a non-linear process governing the relation between different energy bands or the presence of multiple independent variability components of unknown origin \citep{Vaughan_1997coherence}. For small variations, powerlaw pivoting is a linear process and should result in high coherence, but we speculate that variations in coronal geometry such as suggested by \citet{Bollemeijer_2024} may also play a role in decreasing the coherence. In any case, we note that the hard X-ray spectrum of \source{} is complicated and requires multiple Comptonising components \citep{Liu_2024,Yang_2024}, which may also decrease the measured coherence.

\section{Conclusions}

We present a Fourier power- cross- and bi-spectral-timing analysis of BHXRB \source{} in the hard state and HIMS using \nicer{} and \hxmt{} data. Our conclusions can be summarised as follows:
\begin{enumerate}
    \item The QPO waveform, defined by the phase difference $\Psi$ between the QPO fundamental and harmonic frequencies, depends strongly on energy and shows a complex evolution with QPO frequency. For most QPO frequencies, the trend below 15-20 keV is similar to what was observed earlier in other sources by \citet{De_Ruiter_2019} with \rxte, but at higher energies, $\Psi$ evolves in the opposite direction. A possible explanation lies in the pivot energy of the powerlaw at the harmonic frequency being around 15-20 keV during part of the outburst, although more complex models are required to interpret such a claim. QPO phase-resolved spectroscopy can shed more light on the role of the harmonic.
    \item The QPO fundamental and harmonic lags evolve with QPO frequency and energy. At soft X-ray energies covered by \nicer ($<10$ keV), the lags are consistent with those measured for high-inclination sources by \citet{VandenEijnden_2017} and are found to depend strongly on energy. Above $\sim$7 keV, the QPO fundamental lags measured in \hxmt{} data are small and the coherence is high at the QPO frequency, indicating that a single variability component (i.e. the QPO) determines those lags. 
    The lags at the harmonic frequency appear to be dominated by a much broader high frequency bump and depend strongly on energy. We demonstrate how the bi- and cross-spectrum can discern between different variability components at the same Fourier frequency, with the biphase extracting information about the QPO harmonic, while the cross-spectral phase lags at the harmonic frequency are likely due to broadband noise.
    \item The high-energy broadband noise lags evolve strongly during the outburst and can be studied in new detail with the \hxmt{} data set for \source. When the QPO frequency is low, we observe large hard lags at higher frequencies, while for higher QPO frequencies, the source shows large hard lags at frequencies both below and above the QPO frequency. Possibly, these lags are due to power-law spectral pivoting, but detailed modeling is required to study the origin of these large lags.
    \item The QPO waveform and lag behaviour for \source{} are consistent with measurements of BHXRB \sourcem, which is one of the few bright sources with good \hxmt{} coverage in similar accretion states. Both sources spent a relatively large amount of time in the HIMS and are viewed at medium-high inclination ($\sim$40-60$\degree$). The observed similarities show that the complex spectral-timing behaviour of \source{} is not unique and may represent at least a subset of BHXRBs, enabling the study of the innermost X-ray emitting regions in new detail.
\end{enumerate}

\section*{Acknowledgements}
N.B. and this work are supported by the research program Athena with project No. 184.034.002, which is (partially) financed by the Dutch Research Council (NWO). B.Y. is supported by NSFC grants 12322307, 12273026, and 12361131579; by the National Program on Key Research and Development Project 2021YFA0718500; by the Natural Science Foundation of Hubei Province of China 2022CFB167; by the Fundamental Research Funds for the Central Universities 2042022rc0002; Xiaomi Foundation / Xiaomi Young Talents Program. \\
This research makes use of the SciServer science platform (\url{www.sciserver.org}). SciServer is a collaborative research environment for large-scale data-driven science. It is being developed at, and administered by, the Institute for Data Intensive Engineering and Science at Johns Hopkins University. SciServer is funded by the National Science Foundation through the Data Infrastructure Building Blocks (DIBBs) program and others, as well as by the Alfred P. Sloan Foundation and the Gordon and Betty Moore Foundation. \\
This work made use of Astropy:\footnote{\url{http://www.astropy.org}} a community-developed core Python package and an ecosystem of tools and resources for astronomy \citep{astropy:2013, astropy:2018, astropy:2022}. We also made extensive use of NumPy \citep{harris2020array} and SciPy \citep{2020SciPy-NMeth}.

\section*{Data Availability} This research has made use of data obtained through the High Energy Astrophysics Science Archive Research Center Online Service, provided by the NASA/Goddard Space Flight Center. The data underlying this article are available in HEASARC, at \url{https://heasarc.gsfc.nasa.gov/docs/archive.html}. 
This work has made use of the data from the \hxmt{} mission, a project funded by China National Space Administration (CNSA) and the Chinese Academy of Sciences (CAS). All data are public and can be found at \url{http://archive.hxmt.cn/proposal}.
Upon publication, a basic reproduction package for the results and figures presented in this paper will be available on Zenodo at \url{https://doi.org/10.5281/zenodo.14243337}.


\bibliographystyle{mnras}
\bibliography{bibliography} 




\appendix

\section{Light leak in \nicer}
\label{sec:app_lightleak}

The light leak in \nicer, caused by the puncture of a thermal film protecting the detectors from sunlight, affects the data produced in various ways. For an overview of the light leak and its effects, see \url{https://heasarc.gsfc.nasa.gov/docs/nicer/analysis_threads/light-leak-overview/}. In this section, we take a closer look at the effects on the data for \source, especially concerning the timing results. Because the angle between the Sun and \source{} decreased during its 2023 outburst, the effects of the light leak for ISS daytime observations became worse over the weeks. Some of the later observations were made during ISS night, so those are not affected by the light leak.

\begin{figure}
    \centering
    \includegraphics[width=\linewidth]{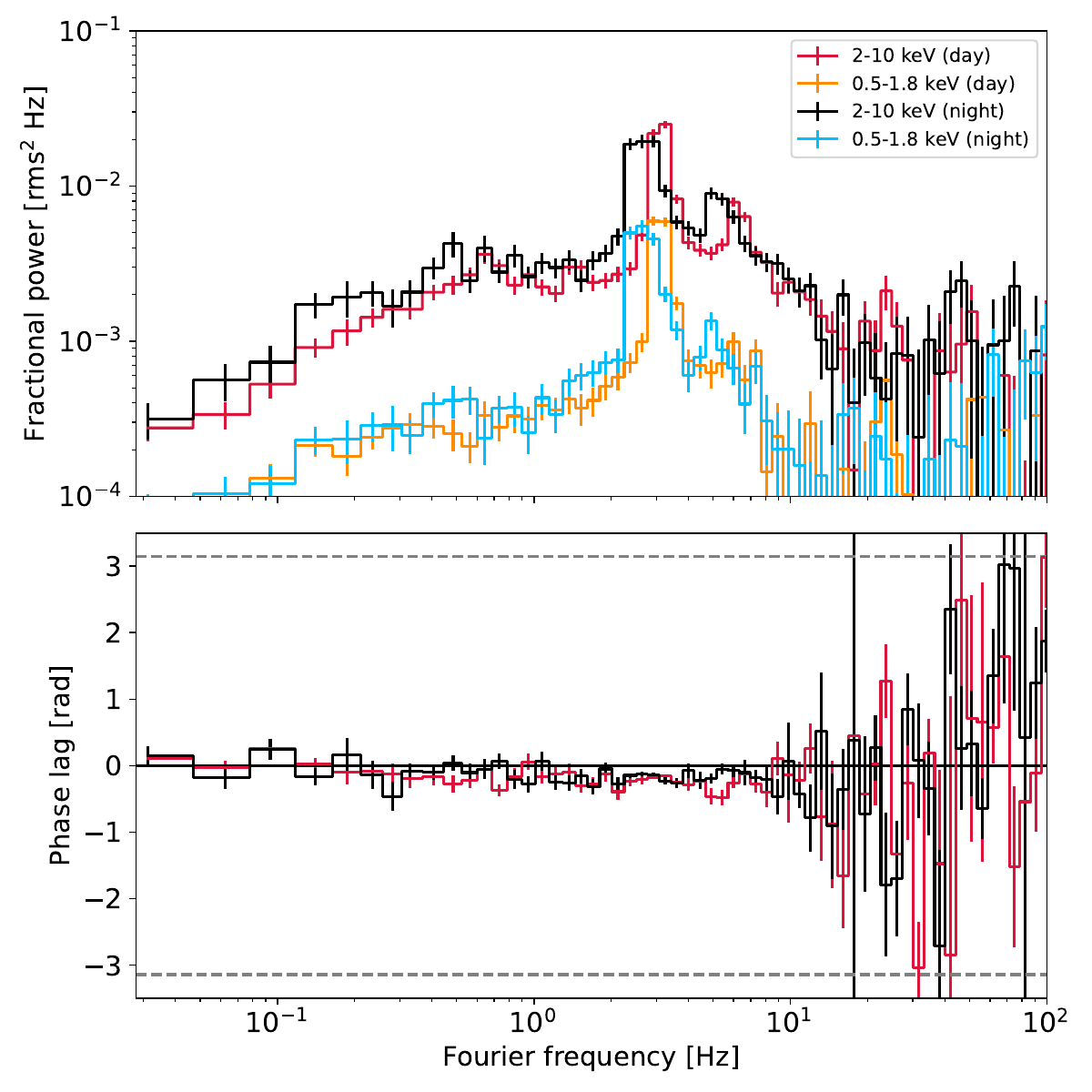}
    \caption{Power- and lag vs frequency spectra for day and night orbits of ObsID 6203980122 of \source. The QPO frequency shifts slightly during the observation, with the nighttime observations corresponding to a lower QPO frequency, which also explains the slightly larger power at lower frequencies for the nighttime data. Otherwise, both the power- and lag vs frequency spectra look similar.}
    \label{fig:timing_comparedaynight}
\end{figure}

One observation, 6203980122, taken on September 16, 2023, has two orbits during ISS night and five orbits during ISS day. Changes in timing properties over the timescale of a few hours are expected to be small, allowing us to study the effects of the light leak in more detail, especially with respect to the spectral-timing. For the purpose of the investigation, we reprocessed 6203980122 with a maximum undershoot rate of 1500 s$^{-1}$ and switched off screening on undershoots setting \texttt{underonlyscr=NONE}. To illustrate the effects of the light leak on timing properties, we made light curves for separate sets of FPMs for the 0.5-1.8 and 2-10 keV energy bands for the two nighttime orbits and two subsequent daytime orbits. In Fig. \ref{fig:timing_comparedaynight}, we show the power spectra and lag vs frequency spectra calculated with these light curves. The QPO frequency shifts during the observation, causing the nighttime observations to have a lower frequency (and slightly broadened) QPO peak, but otherwise, the spectral-timing properties look very similar. 

\begin{figure}
    \centering
    \includegraphics[width=\linewidth]{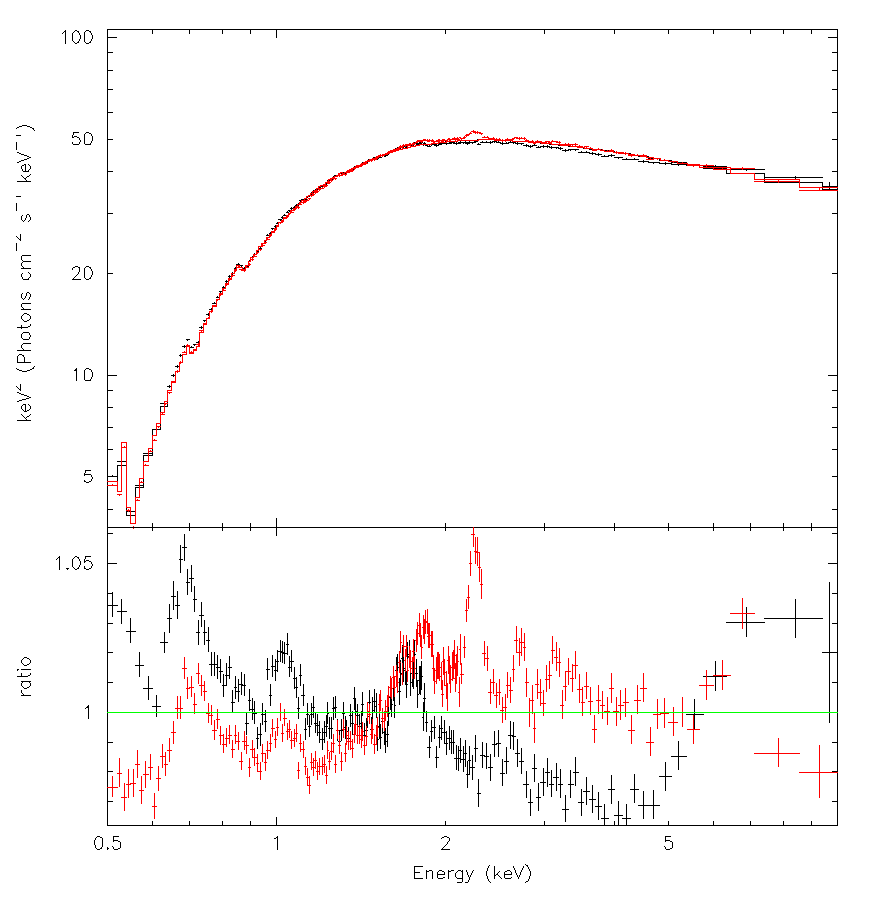}
    \caption{Folded spectra for day- (red) and nighttime (black) observations of ObsID 6203980122. Both spectra were fitted with \texttt{tbabs*nthcomp} with all parameters tied.}
    \label{fig:spectra_comparedaynight}
\end{figure}

We also investigated the effects of the very high undershoot rates on the energy spectrum. We used \texttt{nicerl3-spect} to create the spectra, using the same GTIs as for the light curves. The result is visible in Fig. \ref{fig:spectra_comparedaynight}. To enable comparison between the spectra, we used \textsc{xspec} \citep{Arnaud_1996} and fitted both spectra (red for the daytime, black for the nighttime observations) with \texttt{tbabs*nthcomp} with all parameters tied and only \texttt{nH}, \texttt{Gamma} and \texttt{norm} as free parameters. No systematic errors were added. There are very clear differences between both spectra. First of all, we identify the peak at $\sim$2.2 keV as a strong Gold M edge from the X-ray concentrator gold coating\footnote{\url{https://heasarc.gsfc.nasa.gov/docs/nicer/data_analysis/workshops/NICER-CalStatus-Markwardt-2021.pdf}}. However, it is also clear that the daytime spectrum is harder than the nighttime observation. The change in hardness is unexpected, given the fact that the QPO frequency is slightly higher in the daytime observations, which would correspond to slightly softer spectra, and can probably be attributed to the gain shift caused by the high undershoot rate. Because the precise behaviour of the gain at undershoots $\gtrsim$ 500 s$^{-1}$ is at present still being calibrated, the \nicer{} team does not recommend using the daytime data with very high undershoot rates\footnote{\url{https://heasarc.gsfc.nasa.gov/docs/nicer/analysis_threads/cal-recommend/}}. Future updates to HEASoft may improve the capability to use data under heavy light leak conditions. In any case, spectral-timing studies with reasonably broad energy bands can still provide useful information, as long as their are no or very few packet losses (which affect the light curve shape) and if deadtime effects are mitigated by using the methods described in Section \ref{sec:data_nicer}.

\section{Background estimation for \hxmt{}}
\label{sec:app_background}

\begin{figure}
    \centering
    \includegraphics[width=\linewidth]{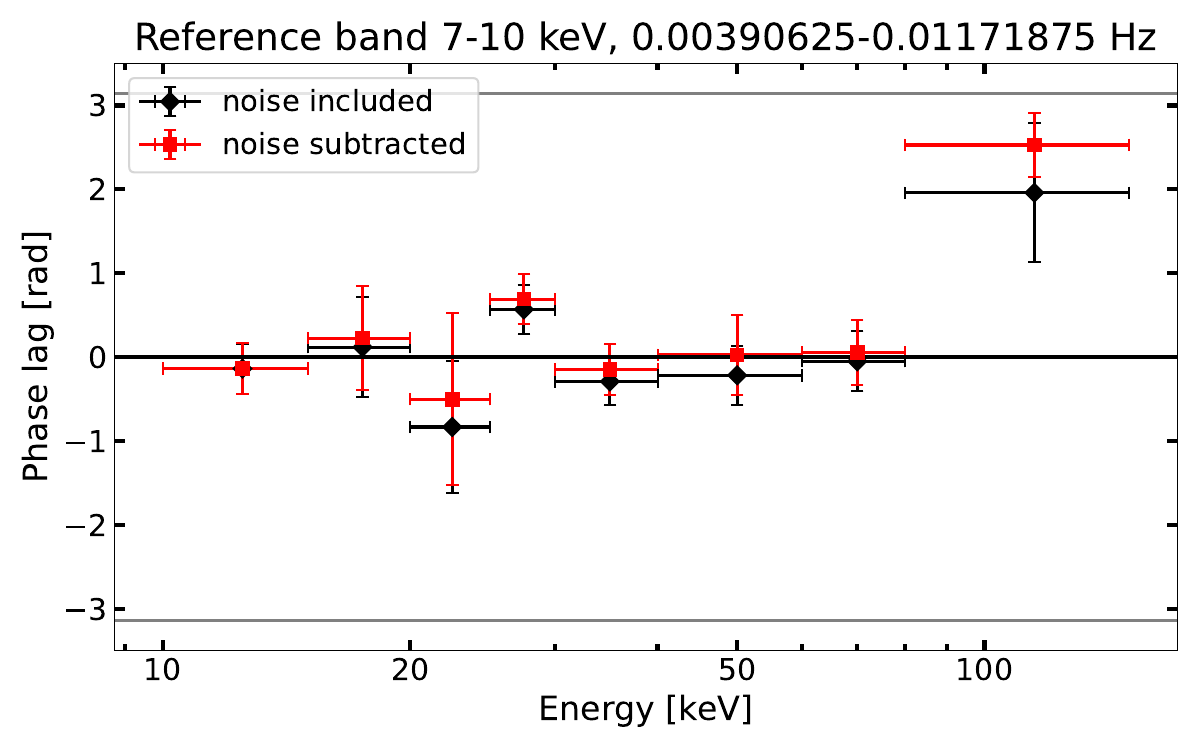}
    \caption{Phase lags at very low frequencies for \hxmt{}  observation P0614338035 with and without background subtracted. The differences are clearly visible, but well within error bars.}
    \label{fig:backgroundeffect}
\end{figure}

We calculated background light curves for the ME and HE instruments for observation P06014338035, which took place on October 5, 2023 (MJD 60222). We chose 9 energy bands in total, 4 for ME and 5 for HE, to investigate the influence of the background on the measured lags. The background light curves were made using the \texttt{hpipeline} tool and the \texttt{mebkgmap} and \texttt{hebkgmap} functionalities in particular. The time resolution of the background light curves is 1 s. We created light curves by making histograms of events by their arrival time and PI with the same time resolution. By calculating the phase lags for these `raw' light curves, we can compare them to light curves from which we subtracted the interpolated background light curves. Because the modelled background varies only on long timescales, we show the phase lags for very low frequencies of 0.00390625-0.01171875 Hz in Fig. \ref{fig:backgroundeffect}, using 256 s segments. The chosen observation and frequency range show the maximum effect of the background, because the source variability is lowest at the end of the HIMS (see e.g. Fig. 1 of \citealt{Yu_2024}) and the background varies only on long timescales. Thus, the effects of background in earlier observations and especially at higher frequencies will be much smaller and we choose to ignore them.

\section{Simulations to test the waveform reconstruction}
\label{sec:app_simulations}

\begin{figure}
    \centering
    \includegraphics[width=\linewidth]{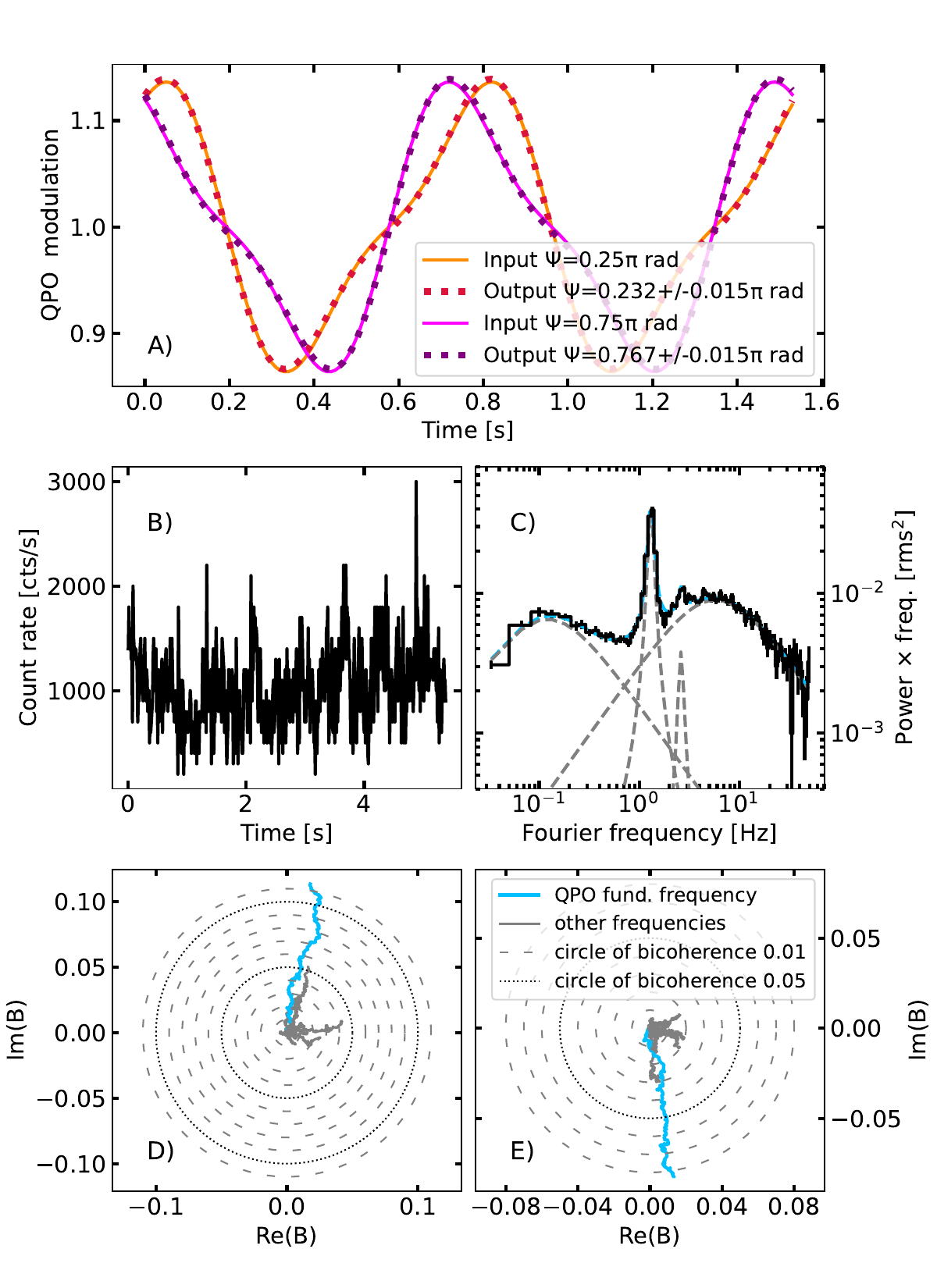}
    \caption{Top: Panel A shows the input simulated waveforms and the output waveforms, reconstructed by calculating the biphase and fitting the power spectrum in panel C. Middle: Panel B shows a short example light curve segment which includes broadband noise and Poisson noise, while panel C shows the simulated power spectrum, fitted with four Lorentzians. Bottom: panels D and E show jellyfish plots similar to those in Fig. \ref{fig:jellyfish}, but with the simulated data. The very good reconstruction of the simulated waveform, as well as the similarity in the jellyfish plots, show that the methods used in Section \ref{sec:qpowaveform} work well.}
    \label{fig:showsims}
\end{figure}

To test our waveform reconstruction techniques, we created simulations that include a quasi-periodic signal with a well-defined waveform in the presence of strong broadband noise. In the methods described in Section \ref{sec:qpowaveform}, we make similar assumptions, which are strengthened by previous research such as the findings of \citet{De_Ruiter_2019}. The broadband noise component of our simulations is based on the methods introduced by \citet{TK95}, which can be used to simulate stochastic variability based on an input power-spectral shape. To simulate the broadband noise, we used two broad Lorentzians based on the fits to the power spectrum of \hxmt{} data with a QPO frequency of around 1.3 Hz, as shown in panel C of Fig. \ref{fig:showsims}. We created a maximally stochastic broadband noise signal with the presented power spectrum following \citet{TK95}. We multiplied the broadband noise by a QPO signal based on the description in \citet{Ingram_2019review} and \citet{Bollemeijer_2024}. The QPO is modelled as the sum of two sinusoids, with one twice the frequency of the other. The sinusoids have a constant phase difference between them, but the phase of the sinusoids has an additional random offset in time. Such phase variations introduce frequency modulation causing the the QPO signal to lose its periodicity and become broader in the power spectrum, while keeping the phase link between the harmonics intact. Although we follow methods very similar to those in Appendix A of \citet{Bollemeijer_2024}, we model the phase variations of the QPO with a low frequency Lorentzian (as prescribed by \citet{Ingram_2019review} on their page 7., Fig. 5) instead of with a random walk. The parameters for the low-frequency Lorentzian signal that modulates the QPO phase are (fractional) amplitude$=0.1$, $\nu_0$=0 and $\sigma=0.1$Hz. Using a random walk to vary the phase of both the fundamental and harmonic leads to the harmonic being a factor $\sim$2 broader than the fundamental, as the harmonic will go out of phase with itself twice as fast as the fundamental. Such broad harmonics are not what we observe in data, where both harmonics have a similar Q-factor. We include Poisson noise in our simulated light curves by renormalizing to an average count rate of 1000 cts/s and drawing the number of counts in each time bin from a Poisson distribution with $\lambda$ being the simulated number of counts in that bin. In total, we simulate two 6000 s light curves, each with their own waveform, and apply our methods for reconstructing the waveform as outlined in Section \ref{sec:qpowaveform}. We refer to our reproduction package for more details on the simulation.

In Fig. \ref{fig:showsims}, we show in top panel A the input and output QPO waveforms, which are very similar, demonstrating that our methods can reconstruct the QPO waveform. We note that in the simulation, the QPO waveform only contains fundamental and harmonic content, while the true QPO waveform in data may be more complicated and require higher harmonics as well. The fact that we do not observe such higher harmonics in power spectra shows that the contribution of these higher harmonics is weak, and to first order, we can parametrize the waveform with two harmonics. The input harmonic amplitude is $0.3A_{\rm{fund}}$, while the amplitude ratio as fitted from the power spectrum is fully consistent with the input at $0.302\pm0.026$. For two input phase differences, $\Psi=0.25\pi$ and $0.75\pi$ rad, we find with the bispectrum that $\Psi=0.232\pm0.015\pi$ and $0.767\pm0.015\pi$ rad. The small shift to the positive real axis, apparent in panels D and E, arises due to the multiplication of the QPO and broadband noise signals, which lead to a positively skewed flux distribution, which has biphase 0. The effect is small, however, and the measured biphase is within 2$\sigma$ of the input biphase. Adding the QPO and broadband noise signals together, instead of multiplying them, does not introduce the flux skewness and leads to measurements of $\Psi$ closer to the input value. We conclude that our methods are well suited to measure the observed evolution and energy-dependence of the QPO waveform.

\section{List of \nicer{} observations}
\label{sec:app_nicerdata}

In Table \ref{tab:nicerdatatable}, we show the list of \nicer{} ObsIDs we used in the current work and the \texttt{nicerl2} settings we used to obtain data. Because changing the settings will affect the quality of the data resulting from the pipeline, we do not recommend using these data for spectroscopy, as explained in Appendix \ref{sec:app_lightleak}. Also, timing results should be interpreted with extra care, e.g. by using separate sets of FPMs to calculate lags.

In all cases, we set \texttt{min\_fpm=1}, ensuring that even observations with very few usable detectors return some data (the default number of detectors required is 7). Also, we use \texttt{thresh\_filter=ALL}, to make sure we include both day and nighttime data. Other changes in reprocessing settings, which are required to obtain data that are affected by the light leak, are listed in Table \ref{tab:nicerdatatable}. For the first four observations, the coordinates of \source{} were not known in enough detail and we allow a slightly larger offset for the source to obtain data. Although doing so can in principle cause timing features,\footnote{\url{https://heasarc.gsfc.nasa.gov/docs/nicer/data_analysis/nicer_analysis_tips.html}} the extra offset is modest in these data and we do not observe instrumental features.

The \texttt{underonly\_range} and \texttt{underonlyscr} parameters regulate the undershoot rates that are allowed per detector. Because of the light leak conditions, some observations have zero exposure time passing the default settings. We applied changes in the \texttt{underonly\_range} parameter, allowing data with up to 1000 undershoots/s in some cases and also tried switching off undershoot screening if \texttt{nicerl2} still returned no events. However, the obtained data had little value, as the number of detectors we could use was generally also very low for those observations. After reprocessing the raw data using \texttt{nicerl2}, we made light curves and excluded any segments with spurious drops in count rate. We refer to our Zenodo package for the code we used for data extraction and further details of the included data.

\begin{table}
    \centering
    \begin{tabular}{c|c|c|c}
        ObsID & Starting date & MJD & Adjusted settings \\
        \hline \hline
        6203980101 & 25-08-2023 & 60181 & \texttt{ang$\_$dist=0.03} \\
        6203980102 & 26-08-2023 & 60182 & \texttt{ang$\_$dist=0.03} \\ 
        6203980103 & 27-08-2023 & 60183 & \texttt{ang$\_$dist=0.03} \\ 
        6203980104 & 28-08-2023 & 60184 & \texttt{ang$\_$dist=0.03} \\ 
        6203980105 & 29-8-2023 & 60185 &  \\
        6703010101 & 29-8-2023 & 60185 & \\
        6203980106 & 30-8-2023 & 60186 & \\
        6750010101 & 30-8-2023 & 60186 & \\
        6750010102 & 31-8-2023 & 60187 & \\
        6203980107 & 31-8-2023 & 60187 & \\
        6203980108 & 1-9-2023 & 60187 & \\
        6703010102 & 1-9-2023 & 60188 & \\
        6203980109 & 2-9-2023 & 60189 & \\
        6750010201 & 3-9-2023 & 60189 & \\
        6203980110 & 3-9-2023 & 60190 & \\
        6203980111 & 4-9-2023 & 60191 & \\
        6703010103 & 4-9-2023 & 60191 & \\
        6203980112 & 5-9-2023 & 60192 & \\
        6750010301 & 6-9-2023 & 60193 & \\
        6203980113 & 6-9-2023 & 60193 & \\
        6203980114 & 7-9-2023 & 60193 & \\
        6703010104 & 7-9-2023 & 60194 & \\
        6750010501 & 7-9-2023 & 60194 & \\
        6750010502 & 8-9-2023 & 60195 & \\
        6203980115 & 8-9-2023 & 60195 & \\
        6203980116 & 9-9-2023 & 60195 & \\
        6203980117 & 11-9-2023 & 60198 & \texttt{underonly\_range="0-800"} \\
        6703010105 & 11-9-2023 & 60198 & \texttt{underonly\_range="0-1000"}\\
        6703010106 & 12-9-2023 & 60199 & \texttt{underonly\_range="0-1000"}\\
        6203980118 & 12-9-2023 & 60199 & \\
        6511080101 & 13-9-2023 & 60200 & \\
        6203980119 & 13-9-2023 & 60200 & \\
        6703010107 & 15-9-2023 & 60202 & \\
        6750010202 & 15-9-2023 & 60202 & \\
        6703010108 & 16-9-2023 & 60203 & \texttt{underonly\_range="0-1000"}\\
        6750010203 & 16-9-2023 & 60203 & \\
        6203980122 & 16-9-2023 & 60203 & \texttt{underonly\_range="0-1000"} \\
        6557020201 & 16-9-2023 & 60203 & \texttt{underonly\_range="0-800"}\\
        6557020202 & 17-9-2023 & 60203 & \texttt{underonly\_range="0-800"}\\
        6203980123 & 17-9-2023 & 60204 & \texttt{underonly\_range="0-1000"}\\
        6203980124 & 18-9-2023 & 60205 & \texttt{underonly\_range="0-1000"}\\
        6703010109 & 18-9-2023 & 60205 & \texttt{underonly\_range="0-1000"}\\
        6203980126 & 20-9-2023 & 60207 & \texttt{underonly\_range="0-1000"}\\
        6203980130 & 2-10-2023 & 60219 & \\
        6203980131 & 3-10-2023 & 60220 & \\
        6703010113 & 3-10-2023 & 60220 & \\
        6703010114 & 4-10-2023 & 60221 & \\
        6557020401 & 4-10-2023 & 60221 & \\
        6557020402 & 5-10-2023 & 60222 & \\
        6203980136 & 9-10-2023 & 60226 & \\
        
    \end{tabular}
    \caption{All \nicer{} observations and their reprocessing settings used in this paper.}
    \label{tab:nicerdatatable}
\end{table}


\bsp	
\label{lastpage}
\end{document}